\documentclass[
reprint,
superscriptaddress,
amsmath,
amssymb,
aps,
prb,
floatfix,
]{revtex4-2}
\usepackage{graphicx}
\usepackage{hyperref}
\usepackage{xcolor}
\usepackage{multirow}

\begin{document}

\title{Integrated \emph{ab initio} modelling of atomic order and magnetic anisotropy for rare-earth-free magnet design: effects of alloying additions in $\mathrm{L}1_0$ FeNi}

\author{Christopher D. Woodgate}
\email{christopher.woodgate@bristol.ac.uk}
\affiliation{Department of Physics, University of Warwick, Coventry, CV4 7AL, United Kingdom}
\affiliation{H. H. Wills Physics Laboratory, University of Bristol, Royal Fort, Bristol, BS8 1TL, United Kingdom}
\author{Laura H. Lewis}
\affiliation{Department of Chemical Engineering, Northeastern University, Boston, MA 02115, USA}
\affiliation{Department of Mechanical and Industrial Engineering, Northeastern University, Boston, MA 02115, USA}
\author{Julie B. Staunton}
\email{J.B.Staunton@warwick.ac.uk}
\affiliation{Department of Physics, University of Warwick, Coventry, CV4 7AL, United Kingdom}

\begin{abstract}
We describe an integrated modelling approach to accelerate the search for novel, single-phase, multicomponent materials with high magnetocrystalline anisotropy (MCA).  For a given system we predict the nature of atomic ordering, its dependence on the magnetic state, and then proceed to describe the consequent MCA, { magnetisation, and magnetic critical temperature (Curie temperature)}.  Crucially, within our modelling framework, the same {\it ab initio} description of a material's electronic structure determines {all} aspects. We demonstrate this holistic method by studying the effects of alloying additions in FeNi, examining systems with the general { stoichiometries Fe$_4$Ni$_3X$ and Fe$_3$Ni$_4X$, for additives} including $X$ =Pt, Pd, Al, and Co. The atomic ordering behaviour predicted on adding these elements, fundamental for determining a material’s MCA, is rich and varied. Equiatomic FeNi has been reported to require ferromagnetic order to establish the tetragonal L1$_0$ order suited for significant MCA. Our results show that when alloying additions are included in this material, annealing in an applied magnetic field and/or below a material’s Curie temperature may also promote tetragonal order, along with an appreciable effect on the predicted { hard magnetic properties}.
\end{abstract}

\date{November 29, 2024}

\maketitle

\section{Introduction}
\label{sec:introduction}

Permanent magnets play a crucial role in modern society, with applications in electrical power generation and conversion, essential in the global transition to clean energy. The `gold-standard' of current permanent magnetic materials are those based on the rare-earth elements, such as Nd$_2$Fe$_{14}$B~\cite{sagawa_new_1984, croat_prfe_1984} and SmCo$_5$~\cite{strnat_family_1967}, which have large magnetic energy products and are widely used in applications~\cite{lewis_perspectives_2013}. However, the rare-earth elements are a constrained resource, with issues around the stability of the supply chain, price volatility, and the environmental impact of their extraction~\cite{mccallum_practical_2014, smith_stegen_heavy_2015, bai_evaluation_2022}. There is therefore a desire to develop new materials which use reduced concentrations of rare-earth elements, or no rare-earth elements at all. In addition, there is currently a `gap' in permanent magnet performance between the rare-earth-based supermagnets and the significantly weaker oxide ferrite magnets~\cite{coey_permanent_2012}.

\begin{figure*}[t]
\includegraphics[width=0.99\linewidth]{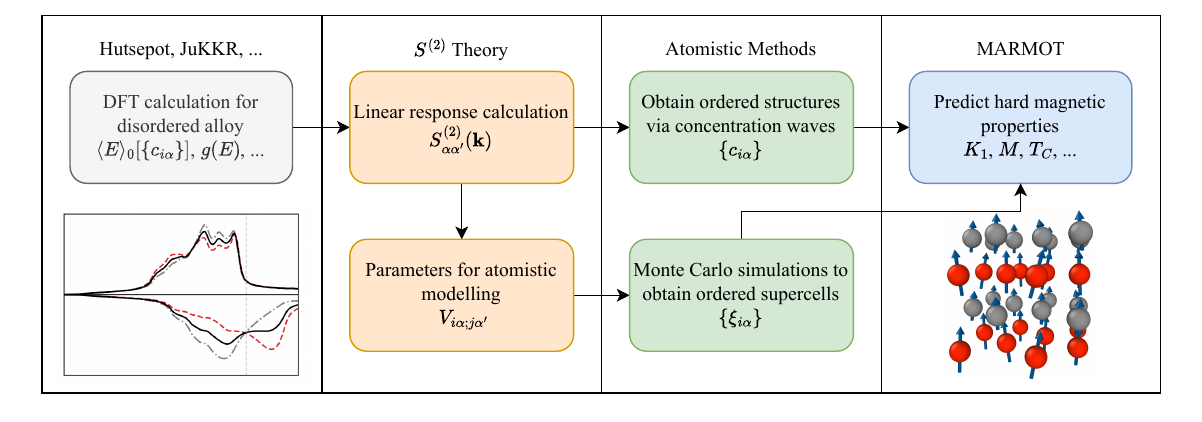}
\caption{Visualisation of the workflow used in this paper for holistically assessing the phase behaviour and subsequent magnetic properties of a given alloy composition.}
\label{fig:workflow}
\end{figure*}

Computational modelling approaches have an important role to play in the process of materials discovery. Not only can they be used to screen potential compositions to suggest which are stable and have the desired physical properties for a particular application, but via an {\it ab initio} description of materials' electronic structure, they also have the potential to advance fundamental physical understanding of why existing materials perform the way they do. This insight can facilitate the optimisation of materials for applications, as well as guide the search for improved compositions and processing techniques.

In this work, we demonstrate a new computational technique which enables holistic assessment of a magnetic alloy's phase behaviour and subsequent magnetocrystalline anisotropy, and is therefore suitable for the discovery of new magnetic materials. As a case study, we apply the approach to suggest new alloy compositions for use as advanced permanent magnets. We begin by describing in detail the binary FeNi system, which is currently under consideration as a potential `gap' magnet~\cite{lewis_inspired_2014}. We then proceed to consider the addition of a third alloying element, studying systems with the composition Fe$_4$Ni$_3X$. The approach is based on complementary techniques for studying the phase behaviour of multicomponent alloys~\cite{khan_statistical_2016, woodgate_compositional_2022, woodgate_short-range_2023, woodgate_interplay_2023, woodgate_competition_2024, woodgate_modelling_2024} and the magnetocrystalline anisotropy of (partially) ordered intermetallic phases~\cite{staunton_temperature_2004, staunton_long-range_2004, staunton_temperature_2006}. Crucially, both the phase behaviour of a material and its magnetocrystalline anisotropy are examined on the same first-principles footing, enabling us to examine how the electronic `glue' of an alloy drives atomic ordering and produces subsequent magnetic properties. This holistic approach is important; it will always be possible to computationally design metastable structures with extraordinary magnetic properties, but for any proposed material to be useful, it must be able to be synthesised in the laboratory.

Our workflow for holistically studying the phase stability and subsequent magnetic properties of a given alloy composition, illustrated in Fig.~\ref{fig:workflow}, proceeds as follows. First, using the Korringa-Kohn-Rostoker (KKR)~\cite{faulkner_multiple_2018} formulation of density functional theory (DFT)~\cite{martin_electronic_2004}, we generate the self-consistent potentials and associated electron density of the disordered alloy---the solid solution---where averaging over substitutional atomic disorder is performed using the coherent potential approximation (CPA)~\cite{soven_coherent-potential_1967}. Then, using the $S^{(2)}$ theory for multicomponent alloys~\cite{khan_statistical_2016, woodgate_compositional_2022, woodgate_short-range_2023, woodgate_interplay_2023, woodgate_competition_2024, woodgate_modelling_2024}, we perform a linear response analysis to rigorously calculate the two-point correlation function, an atomic short-range order (ASRO) parameter, {\it ab initio}. From this linear-response calculation, we can infer (partial) chemical orderings directly using a concentration wave analysis~\cite{khachaturyan_ordering_1978}, or instead fit to a pairwise Hamiltonian describing the internal energy of the alloy that is suitable for atomistic modelling. Once the (partially) ordered phase of interest has been established, we are able to use a DFT-based description of the magnetic torque at both zero and finite temperature to assess the system's magnetocrystalline anisotropy { and magnetic critical (Curie) temperature}~\cite{staunton_temperature_2004, staunton_long-range_2004, staunton_temperature_2006}.

Before presenting the case study of ordered FeNi, it is worth discussing the desirable intrinsic physical properties a material must possess to be suitable for advanced permanent magnetic applications. Putting aside the consideration of shape anisotropy, used to produce magnets such as those of the Alnico family~\cite{skomski_magnetic_2016}, there are three intrinsic physical quantities which determine a material's suitability for use as an advanced permanent magnet. First is the Curie temperature, $T_C$, the temperature below which spontaneous magnetisation occurs even in the absence of an applied magnetic field~\cite{blundell_magnetism_2014}. Second is the saturation magnetisation, $\mathbf{M}_S$, which is the material's magnetic polarization when fully magnetised~\cite{coey_magnetism_2001}. Third is its magnetocrystalline anisotropy (MCA), which measures the energetic cost of rotating the magnetisation vector, $\mathbf{M}$, away from the material's energetically favorable `easy axis' established at the atomic level by spin-orbit coupling and crystal symmetry considerations~\cite{landau_electrodynamics_2009}. The MCA can be related to a material's coercivity, an extrinsic, structure-sensitive property that is a measure of the energetic cost of demagnetising the material~\cite{coey_perspective_2020}. In turn, the coercivity leads to a figure of merit known as the `maximum energy product' which can be used to quantitatively compare the performance different permanent magnets at a given temperature~\cite{coey_perspective_2020}.

Ideally, a permanent magnet will have a high Curie temperature, well above its desired operating temperature, to ensure that its magnetisation remains large during operation. Inevitably, this means that $3d$ transition metals must be incorporated, such as Fe or Co, which have high elemental Curie temperatures of 1043 and 1388 K respectively~\cite{kittel_introduction_2005}. While it is the highly localised $f$ electrons of elements such as Nd and Sm which give rise to the extraordinary magnetic energy products of the rare-earth supermagnets~\cite{patrick_temperature-dependent_2019}, all use significant quantities of Fe or Co in their compositions. This is because these transition metals often stabilise magnetic order at high temperatures and can therefore play a crucial role in determining materials' hard magnetic properties during operation~\cite{bouaziz_crucial_2023}. Similarly, a good permanent magnet will have a large saturation magnetisation, which again requires the use of $3d$ elements such as Fe and Co in any candidate composition.

The MCA of a material is determined by a material's crystal structure and the associated local crystallographic environment around magnetic moments. Permanent magnets generally possess a uniaxial crystal structure~\cite{coey_perspective_2020}, typically tetragonal or hexagonal. These uniaxial crystal structures give rise to uniaxial magnetocrystalline anisotropy, $K$, to leading order, of the form~\cite{landau_electrodynamics_2009}
\begin{equation}
K = K_1 \sin^2 \theta   
\end{equation}
where $\theta$ describes the angle of rotation of the magnetisation away from the easy crystal axis, and $K_1$ is the magnetocrystalline anisotropy constant. In conjunction with a large saturation magnetisation, $\mathbf{M}_S$, a value of $K_1$ larger than approximately 1 MJm$^{-3}$, is considered to be necessary for a material to be of use as a permanent magnet for advanced applications~\cite{coey_permanent_2012}.

In summary, in the search for novel permanent magnets for applications it is necessary to search for a stable intermetallic compound with a uniaxial crystal structure (and associated uniaxial magnetocrystalline anisotropy), a high Curie temperature, and high saturation magnetisation. In addition, with regard to the aforementioned economic and environmental considerations, compositions should be pursued that use reduced concentrations of constrained elements such as Nd, Sm, and Co, as well as noble metals.

\begin{figure}[b]
\includegraphics[width=0.5\textwidth]{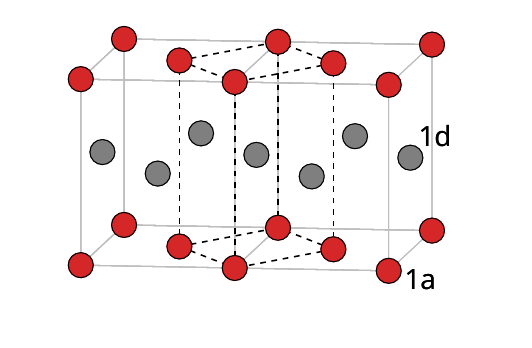}
\caption{Visualisation of $\mathrm{L}1_0$ ordered structure imposed on the face-centred cubic lattice. The conventional simple tetragonal (st) cell is indicated by dashed black lines and contains two non-equivalent sites with Wyckoff labels 1a and 1d. The 1a sites are at the corners of the cell, while the 1d site is at the centre. This st unit cell has lattice parameters $c'=a$, $a'=b'=a/\sqrt{2}$, where $a$ is the conventional fcc lattice parameter. The simple tetragonal representation, with its minimal two-atom basis and a cell elongated in the $\hat{\mathbf{z}}$ direction than in makes the origin of a uniaxial anisotropy in L1\textsubscript{0} materials clear.}
\label{fig:l10_structure}
\end{figure}

A class of materials which fulfill all of the above requirements are those intermetallic compounds forming on the face-centred cubic (fcc) lattice and crystallising in the L1\textsubscript{0} structure, visualised in Fig.~\ref{fig:l10_structure}. This family includes materials already experimentally verified to have superb magnetic properties, such as FePt~\cite{okamoto_chemical-order-dependent_2002, staunton_temperature_2004}, FePd~\cite{shima_lattice_2004, staunton_temperature_2006}, and CoPt~\cite{kanazawa_magnetic_2000, sakuma_first_1994}, but also includes materials consisting entirely of comparatively cheap, earth-abundant elements such as MnAl~\cite{mix_alloying_2017, sakuma_electronic_1994} and FeNi~\cite{neel_magnetic_1964, lewis_inspired_2014, woodgate_revisiting_2023}. In the present work, as a demonstration of our modelling approach, we will focus on the last of these materials, L1$_0$ FeNi.

Although L1\textsubscript{0} FeNi (the meteoritic mineral tetrataenite) does not have a theoretically predicted maximum energy product as large as those of the rare-earth supermagnets~\cite{lewis_inspired_2014}, the relative abundance and availability of its constituent elements, in conjunction with its anticipated production based on principles used in steel processing, mean that its environmental footprint could be significantly lower than for rare-earth-based magnets. In addition, while there remain many open questions concerning tetrataenite's predicted maximum energy product, it is projected to be higher than those of weak magnetic materials such as Alnico and the oxide ferrite magnets~\cite{lewis_inspired_2014, woodgate_revisiting_2023}. As such, tetrataenite may be poised to fill the current gap in  permanent magnetic performance and accessibility. 

However, L1\textsubscript{0} FeNi proves to be extremely challenging to synthesise in a laboratory setting. Although the L1\textsubscript{0} phase is reported to be stable at typical operating temperatures for applications~\cite{lewis_magnete_2014}, its formation is restricted by a low atomic ordering temperature and consequently sluggish kinetics~\cite{neel_magnetic_1964, pauleve_magnetization_1968}. As a result, the fcc solid solution (\textit{Strukturbericht} designation A1) is retained to room temperature, only ordering very slowly under conventional processing techniques. Typically, bulk samples observed experimentally either come from meteorites~\cite{poirier_intrinsic_2015}, which have cooled slowly over a period of millions of years; have had their atomic mobility improved during the annealing process {\it e.g.} by bombardment with high-energy neutrons~\cite{neel_magnetic_1964, pauleve_magnetization_1968}; or have had their structure modified to foster faster diffusion, {\it e.g.} through severe plastic deformation~\cite{montes-arango_discovery_2016}.

\section{Results}
\label{sec:results}

\subsection{FeNi}
\label{sec:feni}

We begin our case study by examining the phase behaviour and magnetocrystalline anisotropy of the equiatomic binary system, FeNi. Using the all-electron HUTSEPOT code~\cite{hoffmann_magnetic_2020}, we perform self-consistent calculations within the KKR formulation~\cite{faulkner_multiple_2018} of DFT\cite{martin_electronic_2004}, using the CPA~\cite{soven_coherent-potential_1967} to average over disorder. This allows us to model the electronic structure of the disordered fcc (A1) phase in both its paramagnetic and ferromagnetic states.

\begin{figure}[b]
\includegraphics[width=0.99\linewidth]{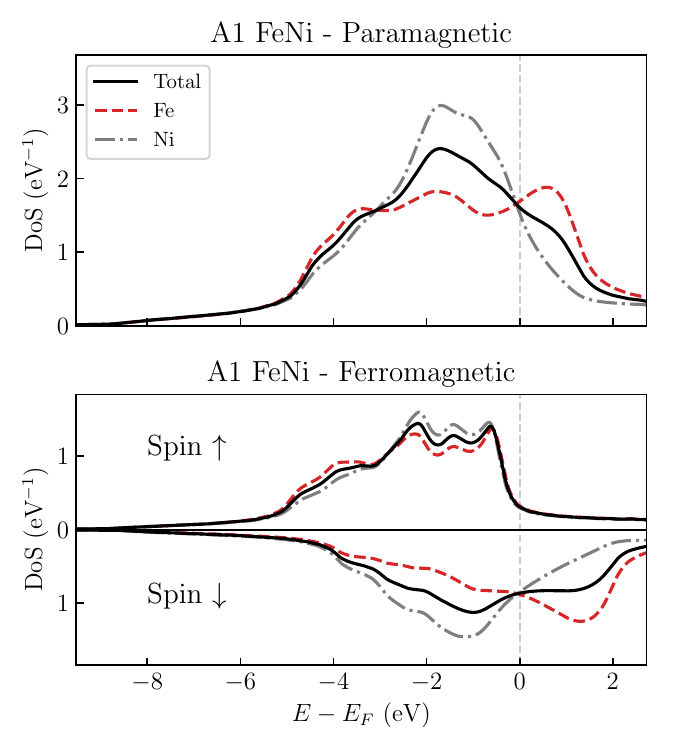}
\caption{Species-resolved density of states for disordered (A1) FeNi in its paramagnetic (upper panel) and ferromagnetic (lower panel) states. The paramagnetic state is modelled within the disordered local moment (DLM) picture. The total DoS curve is given by the average of the species-resolved curves, weighted by their concentrations. For the ferromagnetic state, the upper half of the plot represents the majority-spin DoS, while the bottom half represents the DoS in the minority spin channel. Magnetic order can be seen to clearly alter the nature of the contributions to the total DoS from both Fe and Ni, owing to the  significant difference in exchange splitting of Fe and Ni $d$-states. { For this atomically disordered A1 phase in the ferromagnetic state, the magnetic moments associated with Fe and Ni are 2.53~$\mu_B$ and 0.69~$\mu_B$, respectively. The Curie temperature of A1 FeNi is calculated to be $T_C = 684$~K.}}
\label{fig:feni_dos}
\end{figure}

Fig.~\ref{fig:feni_dos} shows the total and species-resolved density of states for the A1 phase in both its paramagnetic and ferromagnetic states, where the paramagnetic state is modelling within the disordered local moment (DLM) picture~\cite{pindor_disordered_1983, staunton_disordered_1984, gyorffy_first-principles_1985}. The broken magnetic symmetry in the ferromagnetic state can most clearly be seen to differentiate the bonding contributions from majority and minority spin electrons, owing to the different exchange splitting of $d$-states associated with Fe and Ni. This response is expected to significantly alter both the nature of predicted atomic ordering in the system, and the temperature at which this ordering emerges.

{ Again using the disordered local moment (DLM) picture of magnetism at finite temperature as implemented within the MARMOT package~\cite{patrick_marmot_2022} and described in Section~\ref{sec:TC_theory}, we are able to estimate the magnetic critical temperature (\textit{i.e.} Curie temperature) of the disordered solid solution. Above the magnetic critical temperature, the material is paramagnetic, while below it, it is magnetically ordered---ferromagnetic in the case of FeNi. For the atomically disordered, A1 FeNi solid solution, the calculated magnetic critical temperature is $T_C = 684$~K. When the Weiss fields~\cite{patrick_rare-earthtransition-metal_2017}, discussed in Sec.~\ref{sec:TC_theory}, are considered, the normalised eigenvector associated with the magnetic ordering is ($h_\textrm{Fe}$, $h_\textrm{Ni}$) = (0.801, 0.598), indicating that Fe and Ni moments couple ferromagnetically, and that the magnetic ordering is driven primarily by Fe.}

We then use our linear response theory to assess the phase behaviour of this system in both its paramagnetic and ferromagnetic states. The $S^{(2)}$ theory for multicomponent alloys~\cite{khan_statistical_2016, woodgate_compositional_2022, woodgate_short-range_2023, woodgate_interplay_2023, woodgate_competition_2024, woodgate_modelling_2024} has previously been used with success to study atomic arrangements in the Cantor alloy---CrMnFeCoNi---and its derivatives~\cite{woodgate_compositional_2022}, the refractory high-entropy alloys~\cite{woodgate_short-range_2023, woodgate_competition_2024}, and to examine the impact of magnetic ordering on atomic arrangements~\cite{woodgate_interplay_2023}. The theory works with partial atomic occupancies of lattice sites, $\{c_{i\alpha}\}$. For example, for the binary FeNi solid solution, for each site $i$ in the underlying fcc lattice, $c_{i\alpha}=0.5$ for both $\alpha=1$ (Fe) and $\alpha=2$ (Ni). The theory then uses the language of concentration waves~\cite{khachaturyan_ordering_1978} to describe ordered structures, writing
\begin{equation}
c_{i\alpha} = c_\alpha + \sum_{\mathbf{k}} e^{i \mathbf{k} \cdot \mathbf{R}_i} \Delta c_\alpha(\mathbf{k}).
\end{equation}
The overall (average) concentration is represented by $c_\alpha$, while an applied chemical fluctuation (concentration wave) is denoted $\Delta c_\alpha (\mathbf{k})$. For example, the L1$_0$ structure is specified by $\mathbf{k}=(0,0,1)$, in units of $\frac{2\pi}{a}$, and  $\Delta c_{1(2)} = +(-)0.5$ describing alternating layers of Fe and Ni atoms. (For clarity, when tabulating results, we choose to normalise the chemical fluctuation, writing $\Delta \alpha = \Delta c_\alpha / \| \Delta c_\alpha \|$, as this allows the relative size of fluctuations to be objectively compared.) The analysis of the alloy's phase stability then proceeds by performing a Landau-type series expansion of the free energy of the system, allowing the energetic cost of various imposed chemical fluctuations to be assessed. This Landau-type theory allows us to infer the highest temperature $T_\text{ord}$ at which, for some $\mathbf{k}_\text{ord}$, there is a chemical fluctuation $\Delta c_\alpha$ that renders the solid solution unstable, resulting in a chemical ordering. Crucially, the chemical polarisation~\cite{woodgate_compositional_2022}, $\Delta c_\alpha$, informs us how the chemical species partition themselves onto the various sublattices of the predicted ordered structure. The $S^{(2)}$ theory is fully first-principled, and leads to a linear response calculation based on the self-consistent one-electron potentials generated using HUTSEPOT to infer chemical ordering. These potentials can be non-magnetic~\cite{woodgate_short-range_2023, woodgate_competition_2024}, ferromagnetic~\cite{woodgate_interplay_2023}, or paramagnetic~\cite{woodgate_compositional_2022}, as appropriate to the system considered.

Predicted ordering temperatures, wavevectors, and chemical polarisations for FeNi are tabulated in \ref{table:feni_linear_response}. Notably, L$1_0$ ordering, characterised by the wavevector $\mathbf{k}_\text{ord}=(0,0,1)$, is only predicted when we model the material when it is in its ferromagnetic state, reflective of the electronic structure of the material below its Curie temperature. The predicted ordering temperature is 507 K, which compares reasonably well with other theoretical~\cite{tian_density_2019, tian_pressure_2020, izardar_interplay_2020, li_ground-state_2020, li_magnetochemical_2022} estimates and experimental~\cite{neel_magnetic_1964} determinations. These results are highly significant; they suggest that any materials processing aiming to synthesise the L1$_0$ form of FeNi must be carried out below its Curie temperature. It may also be possible that heat treatment of the sample in an applied magnetic field will promote chemical ordering, as applying a magnetic field could induce a (small) spin polarisation to the system~\cite{woodgate_interplay_2023}. 

{ We emphasise that, both for this binary system and the multicomponent systems to follow, the difference in predicted ordered structures and ordering temperatures for difference magnetic states have their origins in the fundamental differences in electronic structure between systems simulated in a paramagnetic state compared to a magnetically ordered one. This is illustrated for the binary FeNi system in Fig.~\ref{fig:feni_dos}.}

\begin{table}[b]
\begin{ruledtabular}
\begin{tabular}{lllrr}
Magnetic State     & $T_\text{ord}$ (K) & $\mathbf{k}_\text{ord}$ & $\Delta$ Fe & $\Delta$ Ni \\ \hline
Paramagnetic       & 175               & $(0, 0.675, 0.675)$    & $0.707$     & $-0.707$    \\
Ferromagnetic      & 507               & $(0, 0, 1)$            & $0.707$     & $-0.707$   
\end{tabular}
\end{ruledtabular}
\caption{Comparison of the predicted atomic ordering temperatures and concentration wave modes for FeNi in both its paramagnetic and ferromagnetic states. Note that we choose to normalise the chemical fluctuation, writing  we choose to normalise the chemical fluctuation, writing $\Delta \alpha = \Delta c_\alpha / \| \Delta c_\alpha \|$. The paramagnetic state is modelled within the disordered local moment picture. It is only when the ferromagnetic state is modelled that an L$1_0$ ordering (indicated by $\mathbf{k}_\text{ord}=(0,0,1)$ is predicted. This suggests that any materials processing aiming to achieve L$1_0$ atomic ordering needs to be performed when the material is below its Curie temperature.}
\label{table:feni_linear_response}
\end{table}

In addition to the above linear response analysis of the phase stability, the $S^{(2)}$ theory also enables extraction of lattice-based atom-atom interaction energies for further atomistic modelling of the phase behaviour of a system. This approach is based on discrete site occupancies, $\{\xi_{i\alpha}\}$, where $\xi_{i\alpha}$=1 if site $i$ is occupied by an atom of species $\alpha$, and $\xi_{i\alpha}$=0 otherwise. The internal energy of the system is described by the conventional Bragg-Williams Hamiltonian~\cite{bragg_effect_1934, bragg_effect_1935}, which takes the form
\begin{equation}
    H(\{\xi_{i\alpha}\}) = \frac{1}{2}\sum_{i \alpha; j\alpha'} V_{i\alpha; j\alpha'} \xi_{i \alpha} \xi_{j \alpha'}.
    \label{eq:b-w}
\end{equation}
The phase behaviour of the system can then be simulated using sampling methods such as the Metropolis Monte Carlo algorithm~\cite{landau_guide_2014}.

For FeNi, our computed atom-atom interaction energies are tabulated in { Supplementary Table 4}. Using these interactions, we perform simulated annealing on an ensemble of 10 simulations of systems of 2048 atoms with periodic boundary conditions applied, and using atomic-swaps rather than substitutions to conserve the overall concentration of each species~\cite{landau_guide_2014}. To assess the nature of atomic ordering in the system, we use conditional probabilities averaged over configurations, denoted $P^{pq}_n$, representing the probability of species $p$ being an $n^\textrm{th}$ nearest-neighbour of species $q$. These parameters therefore represent atomic short-range order (ASRO) parameters. Using the fluctuation-dissipation theorem, it is also possible to estimate the specific heat capacity of the simulated system~\cite{landau_guide_2014}, which helps to identify the temperature at which phase transitions occur.

\begin{figure}[t]
\includegraphics[width=0.99\linewidth]{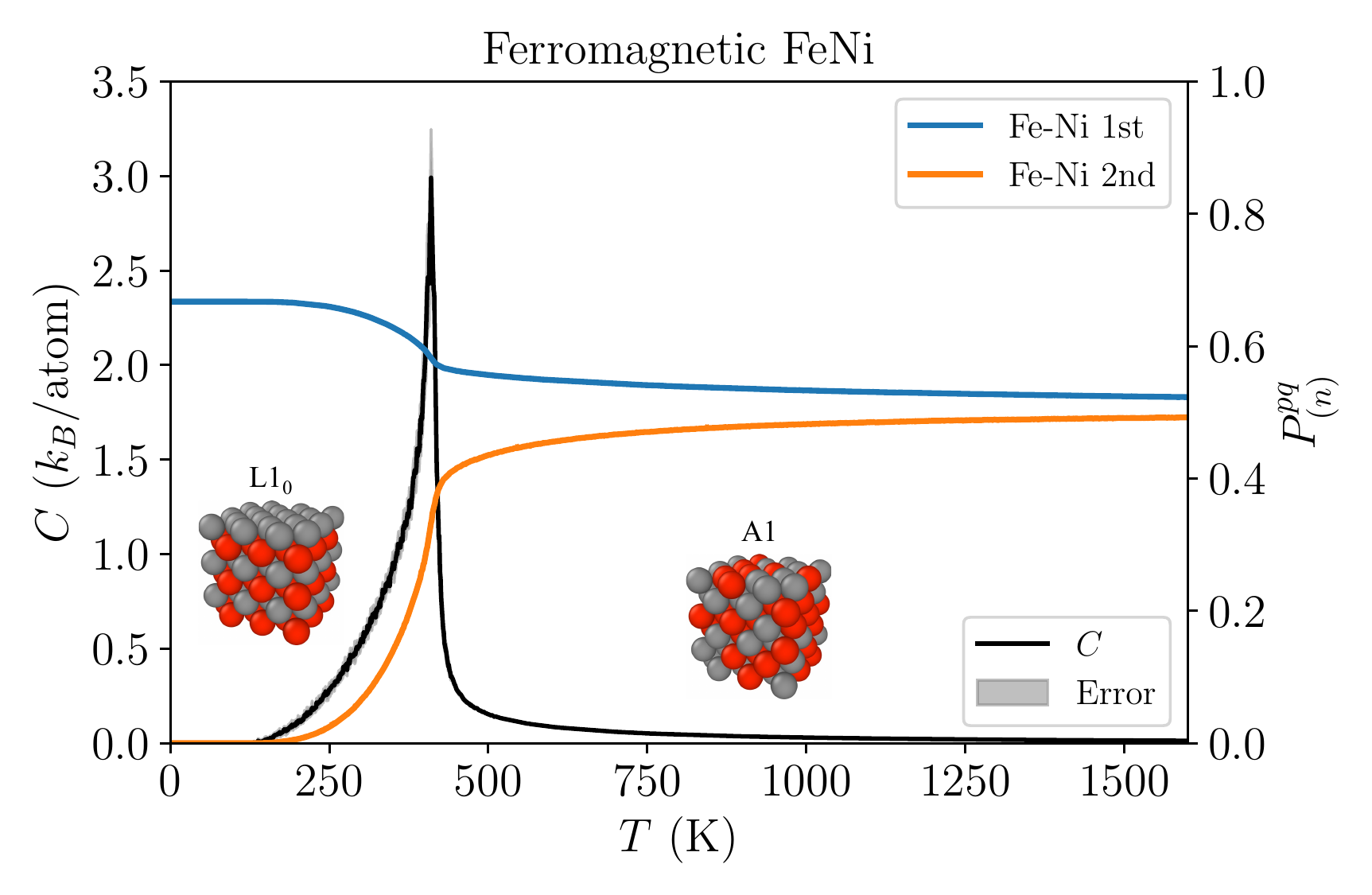}
\caption{Atomic order parameters ($P^{pq}_n$) and specific heat capacity (C) as a function of temperature from lattice-based Monte Carlo simulations for the FeNi system modelled in its ferromagnetic (FM) state. The phase transition just below 500 K is a transition from the atomically disordered fcc (A1) phase to the atomically ordered L1$_0$ phase.}
\label{fig:feni_monte_carlo}
\end{figure}

Visualised in Fig.~\ref{fig:feni_monte_carlo} are our ASRO parameters and specific heat capacity determinations as a function of temperature for FeNi in its ferromagnetic state. The A1-L1$_0$ phase transition is characterised by $P^{\text{Fe-Ni}}_1 = 1/3$, and $P^{\text{Fe-Ni}}_2 = 0$, which occurs a little below 500 K in our simulations, consistent with the result obtained from the linear response analysis, and confirms the L1$_0$ structure as the ground-state of the system.

{
There are a variety of studies reporting contrasting results concerning the significance of vibrational effects when studying the L1$_0$ atomic ordering in FeNi. Some studies have chosen to neglect these effects entirely~\cite{izardar_interplay_2020}, while others have included them in variety of (approximate) ways. Using the cluster variation method, Mohri {\it et al.}~\cite{mohri_first-principles_2009} predicted an L1$_0$ ordering temperature of 523~K without vibrational effects and 483~K with them included, a modest reduction of only 40~K. In contrast, a study by Bonny {\it et al.}~\cite{bonny_feni_2009} employing an embedded atom model (EAM) potential found an L1$_0$ ordering temperature of 1800~K (comparable to the experimental melting temperature of the alloy) without vibrational effects included, and 990~K when they were included, a reduction of 810~K. In between these two extremes lie estimates from a number of other works, including the reduction of 480~K (from 1040~K to 560~K) predicted by Tian \textit{et al.}~\cite{tian_density_2019} using the exact muffin-tin orbital (EMTO) method, the reduction of 280~K (from 920~K to 640~K) predicted by Li and Fu~\cite{li_first-principles_2010} in study using the projector augmented wave (PAW) method, and the reduction of 85~K (from 635~K to 550~K) predicted in a recent study by Ruban~\cite{ruban_qualitative_2024} using the screened generalised perturbation method (SGPM).

Within the modelling approach discussed in this paper, it is not possible to directly include these effects in our lattice-based atomistic modelling. However, for the L1$_0$ ordering predicted for FeNi in its ferromagnetic state, we can estimate the importance of vibrational effects on the ordering by including a vibrational term in our expression for the system's free energy. We employ a Debye treatment of the vibrating lattice as originally proposed by Moruzzi, Janak, and Schwarz~\cite{moruzzi_calculated_1988} to estimate the energy, entropy, and consequent free energy associated with vibrational contributions, which requires an estimate of the Debye temperature of both phases. Details of this calculation are discussed in Sec.~\ref{sec:computational_details}. For the atomically ordered L1$_0$ phase, we obtain a bulk modulus of 189.7~GPa, and an associated Debye temperature of 493~K. For the disordered fcc (A1) phase, we obtain a bulk modulus of 180.2~GPa, and an associated Debye temperature of 430~K. These results confirm the findings of Tian {\it et al.}~\cite{tian_density_2019} that the L1$_0$ phase is elastically stiffer than the disordered (A1) phase. Using our values for the Debye temperatures of the two phases in our expression for the vibrational free energy, we find that our L1$_0$ ordering temperature is reduced from 508~K to 459~K, a reduction of 49~K. If we use the Debye temperatures of Tian \textit{et al.}, the reduction is 59~K, from 508~K to 449~K.

We conclude that vibrational effects do play a role in determining atomic ordering tendencies in Fe-Ni alloys, but that reductions in predicted ordering temperatures are likely to be modest. For the remainder of this work, we choose not to include these effects, but we stress that our predicted ordering temperatures may overestimate those that would be predicted if vibrational effects were fully included.
}

Given a predicted atomically ordered structure, such as L1$_0$ FeNi, we use the same first-principles-based framework to describe its magnetic characteristics and, therefore, suitability for applications. Within our framework, we generate new HUTSEPOT~\cite{hoffmann_magnetic_2020} potentials for the predicted (partially) ordered phases, and then use the MARMOT~\cite{patrick_marmot_2022} code to calculate the phases' magnetic properties~\cite{staunton_temperature_2004}. Our $S^{(2)}$ theory is based on the same underlying formulation of DFT as is used in MARMOT, and therefore we have an integrated description of the electrons driving both chemical ordering and producing the magnetic properties.

We calculate the MCA for perfect L1\textsubscript{0} FeNi at zero-temperature using the same lattice parameters as was employed for the linear response calculation ($c/a = 1$) to be $K_1 = 0.94$~MJm$^{-3}$. Experimental estimates~\cite{neel_magnetic_1964, pauleve_magnetization_1968, shima_structure_2007, mizuguchi_artificial_2011, kojima_l1_2011, kojima_magnetic_2012, kojima_feni_2014, poirier_intrinsic_2015, frisk_resonant_2016, frisk_strain_2017, ito_fabrication_2023} have put the MCA of this material in the region 0.2--0.93~MJm$^{-3}$, while previous theoretical calculations~\cite{izardar_interplay_2020, si_effect_2022, werwinski_ab_2017, miura_origin_2013, tuvshin_first-principles_2021, edstrom_electronic_2014, lewis_magnete_2014, woodgate_revisiting_2023, yamashita_first-principles_2022, yamashita_finite-temperature_2023} have obtained values in the range 0.34--0.96~MJm$^{-3}$. Our magnetisation of $M = 1.66$~$\mu_B$/atom is also consistent with earlier theoretical~\cite{yamashita_first-principles_2022, yamashita_finite-temperature_2023, woodgate_revisiting_2023} and experimental~\cite{frisk_strain_2017} studies. 

{ 
Our theoretical Curie temperature for the L1$_0$ phase of FeNi is 1051~K, again consistent with earlier computational studies~\cite{yamashita_first-principles_2022, yamashita_finite-temperature_2023, woodgate_revisiting_2023}. As for the atomically disordered A1 phase, this ferromagnetic ordering in the L1$_0$ phase is driven by Fe. Considering the Weiss fields (discussed in Sec.~\ref{sec:TC_theory}) for the L1$_0$ phase, the normalised eigenvector associated with the magnetic ordering is \\
($h_\textrm{Fe}$, $h_\textrm{Ni}$) = (0.871, 0.491), confirming that Fe and Ni moments couple ferromagnetically, and that the magnetic ordering is driven primarily by Fe.

The increased Curie temperature for the atomically ordered L1$_0$ phase ($T_C = 1051$~K) compared to the atomically disordered A1 phase ($T_C = 684$~K) is understood to have its origins in the competing magnetic orderings exhibited by Fe constrained to the fcc lattice~\cite{pinski_ferromagnetism_1986}. In particular, the reduced Curie temperature related to the number of Fe-Fe nearest neighbour pairs in the A1 phase compared to the L1$_0$ phase. In the L1$_0$ phase (assuming $c/a=1$) each Fe atom `sees' four Fe atoms an eight Ni atoms at nearest-neighbour distance. But, in the atomically disordered, A1 phase, there are an average of six Fe atoms and six Ni atoms at nearest-neighbour distance. In our modelling, the increased number of Fe-Fe nearest neighbour pairs in the A1 phase compared to the L1$_0$ phase appears to be the main driver in the reduction of the magnetic critical temperature.

We take the above results, and their agreement with existing literature} as verification that our treatment models the L1\textsubscript{0} FeNi system successfully.

\subsection{Fe\textsubscript{4}Ni\textsubscript{3}Pt}

\begin{table}[b]
\begin{ruledtabular}
\begin{tabular}{lllrrr}
Magnetic State     & $T_\text{ord}$ (K) & $\mathbf{k}_\text{ord}$ & $\Delta$ Fe & $\Delta$ Ni & $\Delta$ Pt \\ \hline
Paramagnetic       & 794               & $(0, 0, 1)$            & $0.672$     &  $0.065$    & $-0.737$    \\
Ferromagnetic      & 1163               & $(0, 0, 1)$            & $0.795$     & $-0.237$    & $-0.558$
\end{tabular}
\end{ruledtabular}
\caption{Comparison of the predicted atomic ordering temperatures and concentration wave modes for Fe$_4$Ni$_3$Pt in both its paramagnetic and ferromagnetic states. The paramagnetic state is modelled within the disordered local moment picture. Both orderings are dominated by Pt, on account of the relative atomic size difference between it and the smaller $3d$ elements Fe and Ni, however there is better distinguishment between Fe and Ni in the ferromagnetic state.}
\label{table:fenipt_linear_response}
\end{table}

To promote atomic ordering and enhance magnetocrystalline anisotropy in this class of materials, one approach to consider is the addition of a third alloying element~\cite{tian_alloying_2021}. As an illustrative example of the effects of such alloying additions on atomic arrangements in this material, we consider the addition of a small amount of Pt to the (Fe,Ni) solid solution. Although Pt is unlikely to ever find its way into a mass-produced magnet on account of its high cost and low abundance, it is known that the binary FePt system readily crystallises into the L$1_0$ structure with an atomic ordering temperature of around 1600~K~\cite{von_goldbeck_ironplatinum_1982}. In addition, the L$1_0$  phase of FePt is known to have a large MCA energy ($K_1 \simeq 15$~MJm$^{-3}$) attributed to the heavy $5d$ Pt atoms driving large spin-orbit coupling~\cite{staunton_temperature_2004}. Our expectation is that the addition of Pt to FeNi to form Fe(Ni,Pt) could yield an energetically stable intermetallic phase with magnetocrystalline anisotropy between that of FeNi and FePt. As a demonstration, we consider the substitution of Pt at a concentration of 12.5\%, {\it i.e.} studying a system with the overall composition Fe$_4$Ni$_3$Pt. { The Curie temperature of the A1 Fe$_4$Ni$_3$Pt solid solution is calculated to be $T_C = 715$~K, and its average magnetisation $M=1.61$~$\mu_B$/atom.}

First, in a naive calculation, we calculate the magnetocrystalline anisotropy of an ordered phase of this system assuming that Pt exclusively occupies sites on the Ni sublattice in the L1$_0$ structure. That is, the 1a site is occupied by pure Fe, while the 1d site has chemical occupancy Ni$_{0.75}$Pt$_{0.25}$. Under this assumption, a MCA of $K_1=3.44$~MJm$^{-3}$ is computed, a value significantly larger than that for obtained pure L1$_0$ FeNi. { We also compute a Curie temperature for this ordered phase of $T_C = 1034$~K, comparable to that of L1$_0$ FeNi.} However, these calculations do not account for the phase stability of the compound. To assess the likelihood of formation of the ordered phase, we again apply the $S^{(2)}$ theory for multicomponent alloys~\cite{khan_statistical_2016, woodgate_compositional_2022, woodgate_short-range_2023, woodgate_interplay_2023, woodgate_competition_2024, woodgate_modelling_2024} to this composition.

Tabulated in \ref{table:fenipt_linear_response} are predicted atomic ordering temperatures, concentration wavevectors, and chemical polarisations for Fe$_4$Ni$_3$Pt in both its paramagnetic and ferromagnetic states. In contrast to results obtained for pure FeNi, the wavevector associated with ordering of the Pt-modified composition is found to be $\mathbf{k}_\text{ord}=(0,0,1)$ in both magnetic states. The predicted atomic ordering temperatures are likewise similar, initially suggesting that the magnetic state of the disordered phase might not be as critical to the onset of atomic ordering in the Pt-modified material as in the binary FeNi. However, when the chemical polarisation of the concentration wave is inspected, it can be seen that both the paramagnetic and ferromagnetic orderings are characterised by the arrangements of the Pt atoms, with Fe and Ni distributions largely unaffected and remaining disordered. We associate this dominance with the relative atomic size difference between the $3d$ elements Fe and Ni, and the larger $5d$ element Pt, an effect which has been noted in alloys such as the Ni-Pt system.

From this linear response calculation, we infer an L$1_0$ ordering in both ferromagnetic and paramagnetic states, and at higher temperatures than for the FeNi binary. To predict the (partial) atomic site occupancies, we take the chemical polarisation of the concentration wave and allow this concentration wave to `grow' until (at least) one chemical species has an associated lattice site occupancy of zero, this being the most atomically ordered structure consistent with the concentration wave. For the atomic ordering predicted for the paramagnetic solid solution, the $1a$ site occupancy is given by Fe$_{0.614}$Ni$_{0.386}$, while the $1d$ site occupancy is given by Fe$_{0.386}$Ni$_{0.364}$Pt$_{0.25}$. (See Fig.~\ref{fig:l10_structure} for reference.) The computed {Curie temperature for this ordered state is $T_C=703$~K, while the} MCA for the alloy is quenched to low temperatures is $K_1 = 0.96$~MJm$^{-3}$, a value that is only fractionally larger than that of pure FeNi. However, for the solid solution in a ferromagnetic state, the predicted atomic ordering better separates Fe and Ni, with the $1a$ site occupancy for this condition as Fe$_{0.678}$Ni$_{0.322}$ and the $1d$ site occupancy as Fe$_{0.322}$Ni$_{0.428}$Pt$_{0.25}$. The computed MCA is now higher, at $K_1 = 1.22$~MJm$^{-3}$, a value that is \emph{larger} than the predicted MCA for maximally ordered, unmodified L1$_0$ FeNi. { The computed Curie temperature is also slightly increased, and predicted to be $T_C = 719$~K.}

\begin{figure}[t]
\includegraphics[width=0.99\linewidth]{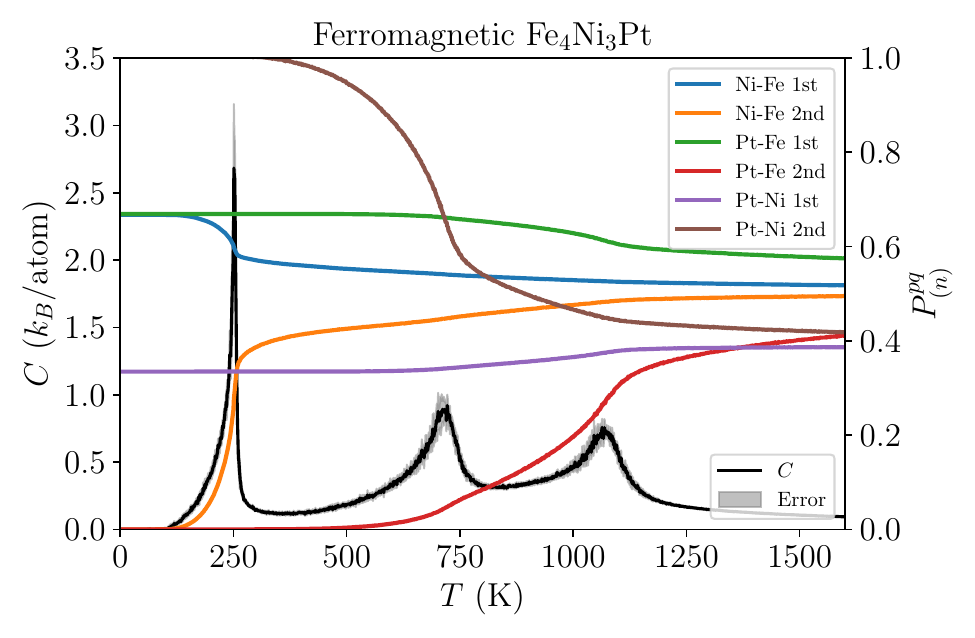}
\caption{Atomic order parameters ($P^{pq}_n$) and specific heat capacity (C) as a function of temperature from lattice-based Monte Carlo simulations for the Fe$_4$Ni$_3$Pt system modelled in its ferromagnetic (FM) state. There are three predicted phase transitions, indicated by the three peaks in the specific heat capacity, with the tetragonal Fe-Ni ordering occurring at the lowest temperature.}
\label{fig:fenipt_monte_carlo}
\end{figure}

\begin{figure}[t]
\includegraphics[width=0.8\linewidth]{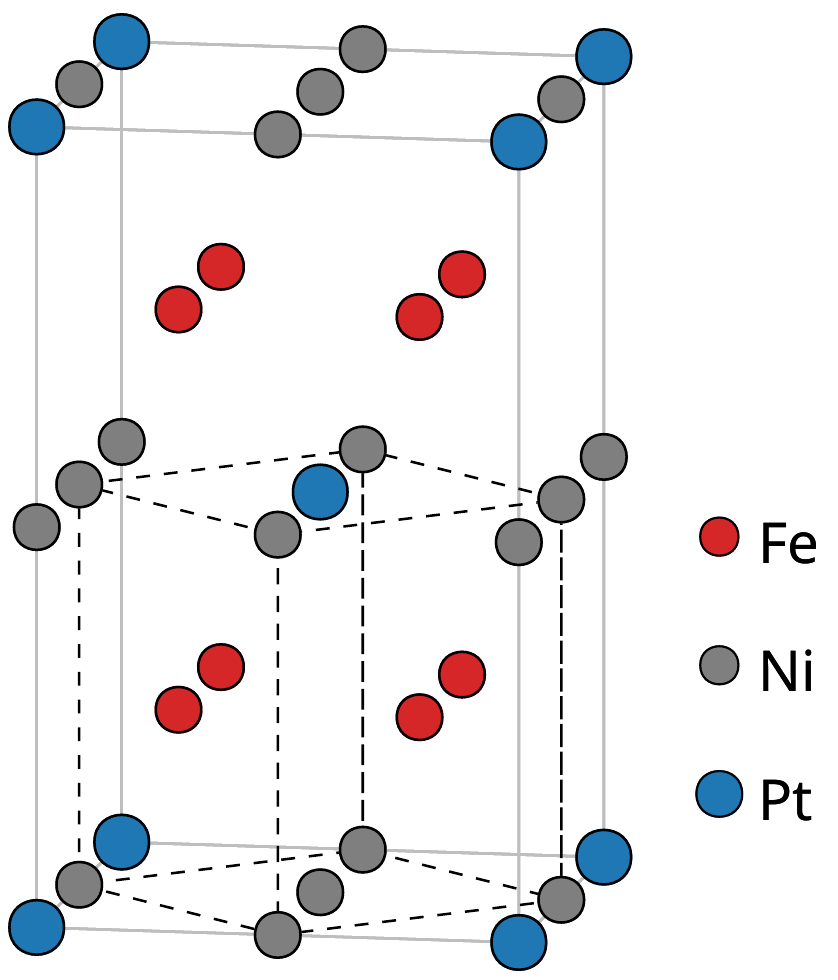}
\caption{The predicted ground state of Fe$_4$Ni$_3$Pt---a tetragonal, atomically ordered, intermetallic phase. The conventional, cubic unit cell of the underlying fcc lattice is marked with black, dashed lines. The MCA of this ordered phase is predicted to be $K_1 = 2.77$ MJm$^{-3}$, approximately three times that of L$1_0$ FeNi. {Its average magnetisation is $M=1.64$~$\mu_B$/atom, and its predicted Curie temperature is $T_C = $985~K.}}
\label{fig:fenipt_ground_state}
\end{figure}

In a multicomponent alloy system such as the Fe$_4$Ni$_3$Pt composition, the linear response calculation may not convey a full picture of the nature of atomic order in the system. However, we are able to recover a simple pairwise atomistic model from it and perform further Monte Carlo simulations to better understand the nature of the atomic arrangements. (See Fig.~\ref{fig:workflow} for an overview of the workflow.) Visualised in Fig.~\ref{fig:fenipt_monte_carlo} are atomic radial distribution probabilities and a measure of the configurational contribution to the specific heat capacity (SHC) of the system. It can be seen that there are in fact three distinct crystallographic orderings which can occur. The first ordering, at relatively high temperature, is cubic and Pt-driven---it represents Pt occupying the corners of the parent fcc lattice, with Ni and Fe arranged in a disordered manner. The second ordering is also Pt-driven, and represents Pt expelling other Pt atoms to third-nearest neighbour occupancies. Finally, the last ordering is an tetragonal ordering of the Fe and Ni atoms, with Pt atoms unaffected. 

The structure of lowest energy for this atomistic model of Fe$_4$Ni$_3$Pt is visualised in Fig.~\ref{fig:fenipt_ground_state}. It has a clear, uniaxial crystal structure, and consists of alternating layers of Fe and ordered Ni-Pt. The computed MCA of this ordered intermetallic phase is $K_1 = 2.77$~MJm$^{-3}$, approximately three times that of equiatomic FeNi. However, within our modelling framework this ground-state structure only becomes thermodynamically stable at low temperatures, suggesting it will be challenging to synthesise it in the laboratory. This low atomic ordering temperature is associated with an increased configurational entropy contribution for a three-component system over that donated by a binary composition, an aspect which will tend to drive down atomic ordering temperatures. In addition, in our simulations, ordering between Pt and other elements at higher temperature `lock' the Pt atoms to particular lattice sites at relatively high temperatures, limiting the availability of lattice sites that allow Fe and Ni to freely interchange positions and therefore inhibiting atomic ordering.

These results, therefore, present a mixed picture. On one hand, adding Pt to the FeNi base composition clearly has the potential to produce a phase with a ground state configuration possessing a high magnetocrystalline anisotropy that is significantly higher than that of binary FeNi. However, the increased configurational entropy and inhibited transformational pathways of the Pt-modified compositon mean that this ground state configuration is only stable at low temperatures with concomitant synthesis challenges.

\subsection{Fe$_4$Ni$_3X$, $X$ = Al, Co, Pd, Mo, Cr}

Building off of the previous example of (FeNi)Pt, to demonstrate the efficacy of our approach for searching for ordered, intermetallic phases with large predicted magnetocrystalline anisotropy energies, and to examine whether alloying provides a viable, cost-effective route for producing atomically ordered L$1_0$ FeNi in the laboratory, we consider five more alloying additions, added individually to the FeNi lattice: $X=$ Al, Co, Pd, Mo, Cr. Al is chosen as a light abundant element known to form intermetallic phases with both Fe and Ni~\cite{sokolovskiy_impact_2022}. Co is chosen as it is known to provide a higher Curie temperature than Fe~\cite{kittel_introduction_2005}, potentially leading to improved magnetic properties. Pd, with an isoelectronic valence structure to Pt~\cite{montes-arango_l10_2015}, is chosen for comparison with the earlier results. Finally, Cr and Mo are chosen as early transition metals, to examine whether the early $3d$ transition metals will have an impact on atomic ordering. { The calculated magnetisation values and magnetic critical temperatures for the A1 solid solution for each of these compositions are given in Table~\ref{table:A1_Fe4Ni3X_magnetism}, while results describing compositional ordering and hard magnetic properties are given in Table~\ref{table:all_additive_results}}.

\begin{table}[b]

\begin{ruledtabular}
\begin{tabular}{lrr}
Composition    & M ($\mu_B$/atom) & $T_C$ (K) \\ \hline
FeNi           & 1.61               & 684       \\
Fe$_4$Ni$_3$Al & 1.32               & 551       \\
Fe$_4$Ni$_3$Cr & 1.08               & 375       \\
Fe$_4$Ni$_3$Co & 1.73               & 835       \\
Fe$_4$Ni$_3$Mo & 1.24               & 309       \\
Fe$_4$Ni$_3$Pd & 1.61               & 755       \\
Fe$_4$Ni$_3$Pt & 1.61               & 715      
\end{tabular}
\end{ruledtabular}
\caption{Average magnetisations ($M$) and calculated magnetic critical temperatures ($T_C$) for the atomically disordered, A1 solid solutions with compositions of the form Fe$_4$Ni$_3X$. Al, Cr, and Mo are all detrimental to the magnetisation and magnetic critical temperature of the solid solution.}\label{table:A1_Fe4Ni3X_magnetism}
\end{table}

\begin{table*}[t]
\begin{ruledtabular}
\begin{tabular}{lcccrrrrrrrrr}
\multirow{2}{*}{Composition} & Magnetic & \multirow{2}{*}{$T_\text{ord}$ (K)} & \multirow{2}{*}{$\mathbf{k}_\text{ord}$ ($2\pi/a$)} & \multirow{2}{*}{$\Delta$Fe} & \multirow{2}{*}{$\Delta$Ni} & \multirow{2}{*}{$\Delta$X} & \multicolumn{2}{l}{$K_1$ (MJm$^{-3}$)} & \multicolumn{2}{l}{$M$ ($\mu_B/$atom)} & \multicolumn{2}{c}{{$T_C$ (K)}} \\ \cline{8-9} \cline{10-11} \cline{12-13}
& State & & & & & & Holistic & Ideal & Holistic & Ideal & {Holistic} & {Ideal} \\ \hline
FeNi & PM & 175 & (0, 0.67, 0.67) & $0.707$ & $-0.707$ & & --- & & --- & & {---} & \\
     & FM & 507 & (0, 0, 1)       & $0.707$ & $-0.707$ & & 0.94 &  0.94  & 1.66 & 1.66 & {1051} & {1051} \\
Fe$_4$Ni$_3$Al & PM &  686 & (0, 0.3, 1) & $0.115$ & $0.643$ & $-0.758$ & --- &  & --- & & {---} &\\
     & FM & 765 & (0, 0.3, 1) & $0.462$ & $0.352$ & $-0.814$ & --- & 1.66 & --- & 1.46 & {---}& {1016} \\
Fe$_4$Ni$_3$Cr & PM &  136 & (0, 0.6, 0.6) & $0.285$ & $-0.805$ & $0.520$ & --- & & --- & & {---}&\\
     & FM & 853 & (0, 0, 1) & $0.790$ & $-0.220$ & $-0.570$ & 0.26 & 1.79 & 1.09 & 1.13 & { 501} & {871}\\
Fe$_4$Ni$_3$Co & PM & 151 & (0, 0, 0) & $0.099$ & $-0.751$ & $0.652$ & --- & & --- & & {---} &\\
     & FM & 433 & (0, 0, 1) & $0.749$ & $-0.656$ & $-0.092$ & 0.42 &  0.31 & 1.75 & 1.75 & { 1055}& {1120} \\
Fe$_4$Ni$_3$Mo & PM & 401 & (0.4, 0.5, 0.6) & $0.169$ & $0.607$ & $-0.776$ & --- & & --- & &  --- &\\
    & FM & 607 & (0, 0.67, 0.67) & $0.344$ & $0.470$ & $-0.814$ & ---  & 1.37 & --- & 1.26 &---&  632 \\
Fe$_4$Ni$_3$Pd & PM & 361 & (0, 0, 1) & $0.460$ & $0.354$ & $-0.814$ & 0.17 & & 1.63 & &  744&\\
    & FM & 696 & (0, 0, 1) & $0.813$ & $-0.338$ & $-0.475$ & 0.20 & 1.05 & 1.64 & 1.66 &  783& 1072 \\
Fe$_4$Ni$_3$Pt & PM & 794 & (0, 0, 1) & $0.672$ & $0.065$ & $-0.737$ &  0.96 & & 1.63 & &  703&\\
    & FM & 1163 & (0, 0, 1) & $0.795$ & $-0.237$ & $-0.558$ & 1.22 & 3.44 & 1.63 & 1.66 &  719 & 1034
\end{tabular}
\end{ruledtabular}
\caption{Computed atomic ordering temperatures ($T_\text{ord}$) and concentration wave modes ($k_\text{ord}$, $\Delta c_\alpha$) for holistic modelling of the preferred ordering induced by an alloying addition  for compositions of the form Fe$_4$Ni$_3X$, followed by the consequent predicted magnetocrystalline anisotropy energies ($K_1$), magnetisations ($M$), { and magnetic critical temperatures ($T_C$) of the ordered phases. An `ideal' value is one where it is assumed that the additive exclusively sits on one sublattice, while a `holistic' value assumes the additive is distributed according to our inferred ordering.} Note that we choose to normalise the chemical fluctuation, writing  we choose to normalise the chemical fluctuation, writing $\Delta \alpha = \Delta c_\alpha / \| \Delta c_\alpha \|$.}
\label{table:all_additive_results}
\end{table*}

Once again, we begin our analysis by calculating an `ideal' MCA, assuming that the elemental addition substitutes only on the Ni sites. Results of these calculations are denoted `ideal' in Table~\ref{table:all_additive_results}. Intriguingly, the addition of Co results in a notable {\it decrease} in MCA compared to that of binary FeNi. This outcome is associated with the Co on the Ni sublattice making a negative contribution to the total magnetic torque in our calculations. The additives Cr, Al, and Mo look promising, as the addition of any of these elements to the Ni sublattice increases the MCA that is predicted by our calculations. It therefore appears that the addition of these elements alters the electronic structure of the material in such a way as to enhance the predicted MCA. This phenomenon has been noted before in other computational studies, {\it e.g.} in a study of general Fe-Ni-Al systems~\cite{sokolovskiy_impact_2022}. Once again, though, we stress that these calculations represent an idealised picture, and the thermodynamic stability of a given phase must be assessed before drawing conclusions as to its suitability as a gap permanent magnet.

To this end, results of the chemical stability analysis for the system Fe$_4$Ni$_3X$, for the considered additions, $X =$ Al, Co, Pd, Mo, Cr, are tabulated in \ref{table:all_additive_results}. The chemical stability analysis is always performed for the system modelled in both its paramagnetic and ferromagnetic states, to assess the degree to which chemical order is coupled to the magnetic state. Where an L1\textsubscript{0} ordering is inferred, indicated by $\mathbf{k}_\text{ord} = (0, 0, 1)$, the maximally ordered L1\textsubscript{0} structure that is consistent with this type of ordering is derived, and subsequently its MCA is calculated. Our results for predicted chemical ordering demonstrate that the assumption that the additive will substitute for Ni is flawed; our inferred chemical ordering never leads to a perfect L1\textsubscript{0}-like structure and, therefore, the anisotropy is always reduced compared to that obtained from the `ideal' lattice structure. This result demonstrates that it is crucial to simultaneously examine the phase behaviour of a candidate material and quantify its magnetic properties, to ensure that any proposed intermetallic phase will be thermodynamically stable, and thus realizable.

The computed outcomes provided in Table~\ref{table:all_additive_results} are examined one by one. Considering first the results for the addition of Al, we see that, for both the paramagnetic and ferromagnetic cases, the concentration wavevector characterising predicted ordering is not commensurate with a simple ordered structure. This outcome is attributed to competing Ni-Al and Fe-Al interactions. This reasoning is supported by experimental finding: the Ni-Al system has a strong tendency to form an intermetallic phase on the fcc lattice~\cite{okamoto_-ni_2004}, while the Fe-Al system forms on a bcc lattice~\cite{von_goldbeck_ironaluminium_1982}. In our Monte Carlo simulations, the results of which are visualised in { Supplementary Figure 6}, the (FeNi)Al system tends to phase segregate into Al-rich and Al-deficient regions. Next considering the results for Cr and Co additions, in both cases an L1$_0$ ordering is predicted only in the case of the system being in the magnetically ordered state while, in contrast, L1$_0$ ordering is not predicted for the paramagnetic state. For Mo additions, L1$_0$ ordering is not predicted in either magnetic state. Finally, the results for Pd additions are comparable with those for the addition of Pt, although the ordering temperatures and the MCA values are notably reduced. The decreased ordering temperature is associated with a reduced difference in atomic radius between chemical species, in turn reducing chemical ordering tendencies. The reduced MCA is associated with a reduced spin-orbit coupling in the system, associated with the smaller mass of the Pd ions compared to Pt ions.

\begin{table}[b]

\begin{ruledtabular}
\begin{tabular}{lrr}
Composition    & M ($\mu_B$/atom) & $T_C$ (K) \\ \hline
FeNi           & 1.61               & 684       \\
Fe$_3$Ni$_4$Al & 1.05               & 638       \\
Fe$_3$Ni$_4$Cr & 0.89               & 391       \\
Fe$_3$Ni$_4$Co & 1.49               & 850       \\
Fe$_3$Ni$_4$Mo & 1.03               & 327       \\
Fe$_3$Ni$_4$Pd & 1.36               & 725       \\
Fe$_3$Ni$_4$Pt & 1.37               & 690      
\end{tabular}
\end{ruledtabular}
\caption{Average magnetisations ($M$) and calculated magnetic critical temperatures ($T_C$) for the atomically disordered, A1 solid solutions with compositions of the form Fe$_3$Ni$_4X$. Al, Cr, and Mo are all detrimental to the magnetic critical temperature of the solid solution.}
\label{table:A1_Fe3Ni4X_magnetism}
\end{table}

\begin{table*}[t]
\begin{ruledtabular}
\begin{tabular}{lcccrrrrrrrrr}
\multirow{2}{*}{Composition} & Magnetic & \multirow{2}{*}{$T_\text{ord}$ (K)} & \multirow{2}{*}{$\mathbf{k}_\text{ord}$ ($2\pi/a$)} & \multirow{2}{*}{$\Delta$Fe} & \multirow{2}{*}{$\Delta$Ni} & \multirow{2}{*}{$\Delta$X} & \multicolumn{2}{l}{$K_1$ (MJm$^{-3}$)} & \multicolumn{2}{l}{$M$ ($\mu_B/$atom)} & \multicolumn{2}{c}{{$T_C$ (K)}} \\ \cline{8-9} \cline{10-11} \cline{12-13}
& State & & & & & & Holistic & Ideal & Holistic & Ideal & {Holistic} & {Ideal} \\ \hline
FeNi & PM & 175 & (0, 0.67, 0.67) & $0.707$ & $-0.707$ & & --- & & --- & & {---} & \\
     & FM & 507 & (0, 0, 1)       & $0.707$ & $-0.707$ & & 0.94 &  0.94  & 1.66 & 1.66 & {1051} & {1051} \\
Fe$_3$Ni$_4$Al  & PM & 866 & (0, 0, 1)         & $ 0.159$ & $-0.773$ & $ 0.614$ & 0.04 & & 1.19 & & 673 &\\
                & FM & 895 & (0, 0.2, 1)       & $ 0.089$ & $ 0.658$ & $-0.748$ & ---  & 0.93 & --- & 1.24 &  ---&  841 \\
Fe$_3$Ni$_4$Cr  & PM & 213 & (0, 0.6, 0.6)     & $ 0.341$ & $-0.813$ & $ 0.472$ & ---  & & --- & &  ---&\\
                & FM & 694 & (0, 0, 1)         & $ 0.773$ & $-0.159$ & $-0.614$ & 0.22 & 0.36 & 0.87 & 1.06&  501 &  625\\
Fe$_3$Ni$_4$Co  & PM & 212 & (0, 0.75, 0.75)   & $ 0.774$ & $-0.612$ & $-0.162$ & ---  & & --- & &  ---&\\
                & FM & 531 & (0, 0, 1)         & $ 0.737$ & $-0.673$ & $-0.063$ & 0.48 & 1.16 & 1.53 & 1.57&  1003 &  1046\\
Fe$_3$Ni$_4$Mo  & PM & 468 & (0, 0.67, 0.67) & $-0.070$ & $ 0.739$ & $-0.670$ & ---  & & --- & &  --- &\\
                & FM & 727 & (0, 0.67, 0.67) & $ 0.138$ & $ 0.628$ & $-0.766$ & ---  & 0.65 & --- & 1.17 &  --- &  632\\
Fe$_3$Ni$_4$Pd  & PM & 373 & (0, 0, 1)         & $ 0.159$ & $-0.773$ & $ 0.614$ & 0.01 & & 1.38 & &  713&\\
                & FM & 688 & (0, 0, 1)         & $ 0.816$ & $-0.436$ & $-0.380$ & 0.22 & 0.55 & 1.40 & 1.41 &  780 &  882\\
Fe$_3$Ni$_4$Pt  & PM & 639 & (0, 0, 1)         & $ 0.757$ & $-0.114$ & $-0.643$ & 0.84 & & 1.39 & &  687 &\\
                & FM & 875 & (0, 0, 1)         & $ 0.814$ & $-0.356$ & $-0.459$ & 0.85 & 0.00 & 1.39 & 1.43 &  716 &  848\\
\end{tabular}
\end{ruledtabular}
\caption{Computed atomic ordering temperatures ($T_\text{ord}$) and concentration wave modes ($k_\text{ord}$, $\Delta c_\alpha$) for holistic modelling of the preferred ordering induced by an alloying addition for compositions of the form Fe$_3$Ni$_4X$, followed by the consequent predicted magnetocrystalline anisotropy energies ($K_1$), magnetisations ($M$), { and magnetic critical temperatures ($T_C$) of the ordered phases. An `ideal' value is one where it is assumed that the additive exclusively sits on one sublattice, while a `holistic' value assumes the additive is distributed according to our inferred ordering.} Note that we choose to normalise the chemical fluctuation, writing  we choose to normalise the chemical fluctuation, writing $\Delta \alpha = \Delta c_\alpha / \| \Delta c_\alpha \|$.}
\label{table:Fe3Ni4X_all_additive_results}
\end{table*}

{ We note that our prediction that the addition of Cr, and Co (modelled in the ferromagnetic state) increase the predicted ordering temperature in an Fe-rich system is in reasonable agreement with Ref.~\cite{tian_alloying_2021}. However, our predicted increased ordering temperature upon the addition of Al is in disagreement; Ref.~\cite{tian_alloying_2021} predicts a decreased ordering temperature except at very low Al concentrations. We suggest that this discrepancy has its origins in the differences between our modelling approach and that used in Ref.~\cite{tian_alloying_2021}. These authors model a system where additives are fixed and distributed homogeneously across all lattice sites, while Fe and Ni atoms are allowed to order. This is different to our approach, where we allow \textit{all} partial lattice site occupancies to vary according to the thermodynamics.}

{
\subsection{Fe$_3$Ni$_4X$, $X$ = Al, Co, Pd, Mo, Cr}

The above results consider the case where we replace a small amount of Ni in the composition with an additive, $X$. This reasoning for this is that, as outlined in Sec.~\ref{sec:introduction}, it is necessary that any candidate permanent magnet have a large magnetisation at room temperature to ensure a sufficiently large magnetic energy product. However, it may be the case that substituting the additive for Fe may promote stronger ordering tendencies. To address this aspect, we now consider compositions of the form Fe$_3$Ni$_4X$. { The calculated magnetisation values and magnetic critical temperatures for the A1 solid solution for each of these compositions are given in Table~\ref{table:A1_Fe3Ni4X_magnetism}, while results describing compositional ordering and hard magnetic properties are given in Table~\ref{table:Fe3Ni4X_all_additive_results}}.

As expected, for the ordered phases, we see that magnetisation values are typically reduced for compositions of the form Fe$_3$Ni$_4X$ compared to those of the form Fe$_4$Ni$_3X$. This is because Fe has a substantially larger magnetic moment associated with it than Ni. (For example, in L1$_0$ FeNi, we find that the moment associated with Fe is 2.66~$\mu_B$, while the moment associated with Ni is 0.65~$\mu_B$.) We note that our calculated `ideal' anisotropies, \textit{i.e.} those where we fix one sublattice to be purely occupied by Ni and the other is given occupancy Fe$_{0.75}X_{0.25}$, are altered compared to the results of Table~\ref{table:all_additive_results}. For example, for $X=$ Cr, Pt, our predicted value of $K_1$ is substantially reduced, while for $X=$Co, it is significantly enhanced. These findings therefore emphasise the strong connection between atomic arrangements and magnetic properties in this class of system.

When the `holistic' results are considered, we find that our predicted orderings for the Ni-rich compositions are broadly consistent with those obtained for the Fe-rich compositions above, with the exceptions being the additions of Al, Co, and Pd when these additives are modelled with the system in a paramagnetic state. For the addition of Pd, the calculation suggests layering of Ni with disordered Fe-Pd. However, the predicted ordering temperature (373~K) is still too low to be of experimental interest. For the addition of Co, we predict a marginally different ordering than for the Fe-rich composition but, again, at such a low temperature that it is not of interest. Finally, for the addition of Al, we now have a wavevector classifying ordering of $\mathbf{k}_\text{ord}= $ (0, 0, 1). However, this ordering is dominated by Ni and Al and, as for our earlier results, we attribute this to the fact that the Ni-Al system has a strong tendency to form an intermetallic phase on the fcc lattice~\cite{okamoto_-ni_2004}. That the addition of Co increases the atomic ordering temperature (in the ferromagnetic case) is in alignment with Ref.~\cite{tian_alloying_2021}, but these authors report that Cr and Al decrease their predicted atomic ordering temperature, a result that differs from ours. As for the Fe-rich results, we emphasise that these differences can, in part, be attributed to the differences between the two modelling approaches.
}

\section{Discussion}
\label{sec:conclusions}

There are three key conclusions that can be drawn from this study. The first is that, when considering the thermodynamic stability of a candidate hard magnetic phase, it is \emph{crucial} to consider its magnetic state as well. Our results for FeNi suggest that L1$_0$ ordering is \emph{only} predicted once the parent system has ferromagnetically ordered, suggesting that processing of this phase should take place below its Curie temperature and/or in an applied magnetic field, thereby promoting desired atomic ordering~\cite{lewis_accelerating_2023}. A similar story holds true when considering alloying additions to this system; different treatments of the magnetic state within our modelling can result in different predicted atomic ordering tendencies.

The second key conclusion is that it is vital to assess the thermodynamic stability of any proposed hard magnetic compound. It will always be possible to imagine metastable crystal structures with extraordinary magnetic properties, but it is only by considering phase equilibria, as is conducted in this study, that a holistic assessment can be made about a candidate material's suitability for applications.

Finally, although these results, and in particular results presented for Pt additives in FeNi, suggest that some alloying additions may yield increased ordering temperatures and crystal structures with improved magnetocrystalline anisotropy compared to the base FeNi binary, it is not the case that all additives promote L1$_0$ ordering. It is necessary, therefore, to perform a complete stability analysis of a given specified composition, rather than assume attainment of an L1$_0$ ordered phase from the outset.

Further work should seek to perform high-throughput searches across a wider range of potential compositions, to search for novel, multi-component alloy phases with desirable magnetic properties. Results of these computational studies can be implemented into experimental activities, to more rapidly realize new types of advanced magnetic materials. We are in the process of adapting our codes for such studies.

\section{Methods}
\label{sec:theory}

\subsection{The $\mathbf{S^{(2)}}$ theory for multicomponent alloys}
\label{sec:chemical_ordering}

Our approach for modelling atomic order in multicomponent alloys uses a Landau-type expansion of the free energy of the system about a homogeneous (disordered) reference state to obtain the two-point correlation function, an ASRO parameter, {\it ab initio}. The effects of the response of the electronic structure and the rearrangement of charge due to the applied inhomogeneous chemical perturbation are fully included. The theory is the natural multicomponent extension of the $S^{(2)}$ theory developed for binary alloys~\cite{gyorffy_concentration_1983, staunton_compositional_1994}, and it has its groundings in statistical physics and the in seminal papers on concentration waves authored by Khachaturyan~\cite{khachaturyan_ordering_1978} and Gyorffy and Stocks~\cite{gyorffy_concentration_1983}. Full details of the theory, its implementation, and extensive discussion can be found in earlier works~\cite{khan_statistical_2016, woodgate_compositional_2022, woodgate_short-range_2023, woodgate_interplay_2023, woodgate_competition_2024, woodgate_modelling_2024}.

The approach assumes a fixed, ideal lattice, face-centred cubic (fcc) for the alloys studied in this paper, which represents the averaged atomic positions in the solid solution. A description of the atomic arrangements in a substitutional alloy with this fixed underlying lattice is given by the site occupation numbers, $\{\xi_{i\alpha}\}$, where $\xi_{i\alpha}$=1 if site $i$ is occupied by an atom of species $\alpha$, and $\xi_{i\alpha}$=0 otherwise. Each lattice site must be constrained to have one (and only one) atom sitting on it, expressed as $\sum_\alpha \xi_{i\alpha}$=1 for all lattice sites $i$. The overall concentration of a chemical species, $c_\alpha$ is given by $c_\alpha = \frac{1}{N} \sum_i \xi_{i\alpha}$, where $N$ is the total number of lattice sites in the system. A natural choice of atomic long-range order parameter is given by the ensemble average of the site occupancies, $c_{i\alpha} = \langle \xi_{i\alpha} \rangle$, where the $\{c_{i\alpha}\}$ are referred to as the site-wise concentrations. In the atomically disordered limit, corresponding to the solid solution at high temperatures, these occupancies will be spatially homogeneous and take the values of the overall concentration of each chemical species, {\it i.e.} $\lim_{T \to \infty} c_{i\alpha} = c_\alpha$. Below any atomic disorder-order transition, however, the site occupancies acquire a spatial dependence. These spatially dependent occupancies (or concentrations) can be written as a fluctuation to the concentration distribution of the homogeneous system, $c_{i\alpha} = c_\alpha + \Delta c_{i\alpha}$. As the underlying system is lattice-based and possesses translational symmetry, it is natural to write these fluctuations in reciprocal space using a concentration wave formalism, as pioneered by Khachaturayan\cite{khachaturyan_ordering_1978}. In this manner we write
\begin{equation}
c_{i\alpha} = c_\alpha + \sum_{\mathbf{k}} e^{i \mathbf{k} \cdot \mathbf{R}_i} \Delta c_\alpha(\mathbf{k})
\end{equation}
to describe a chemical fluctuation, where $\mathbf{R}_i$ is the lattice vector with corresponding occupancy $c_{i\alpha}$. As an illustrative example, Fig.~\ref{fig:l10_structure} shows L$1_0$-type crystallographic order imposed on the fcc lattice for an equiatomic, $AB$ binary system, $c_\alpha = (\frac{1}{2}, \frac{1}{2})$. The completely L$1_0$-ordered structure, represented by alternate layers of atoms of species $A$ and $B$, is described by $\mathbf{k} = \{(0,0,1), (0,0,-1)\}$, with the (normalised) change in concentration $\Delta c_\alpha = \frac{1}{\sqrt{2}} (-1, 1)$. This formalism is extensible to multicomponent alloys.

Correlations between atomic species above any disorder-order transition temperature are quantified by the two-point correlation function, written as \begin{equation}
\Psi_{i\alpha j \alpha'} = \langle \xi_{i\alpha}\xi_{j\alpha'} \rangle - \langle \xi_{i\alpha} \rangle \langle \xi_{j\alpha'}\rangle,
\end{equation}
which is an atomic short-range order (ASRO) parameter. This quantity is intrinsically related to the energetic cost of chemical fluctuations~\cite{khan_statistical_2016}, as ASRO will be dominated by those chemical fluctuations which are least costly energetically.

To find the energy cost we approximate the free energy, $\Omega$, of an alloy with an inhomogeneous site-wise concentration distribution, $\{c_{i\alpha}\}$, by
\begin{equation}
\begin{aligned}
    \Omega^{(1)}[\{c_{i\alpha}\}] =& -\frac{1}{\beta} \sum_{i\alpha} c_{i\alpha} \ln c_{i\alpha} \\
    &- \sum_{i\alpha} \nu_{i\alpha} c_{i\alpha} + \langle \Omega_\text{el} \rangle_0 [\{c_{i\alpha}\}],
    \label{eq:free_energy}
\end{aligned}
\end{equation}
where the three terms on the right-hand side of Eq. (2) describe entropic contributions, site-wise chemical potentials, and an average of the electronic contribution to the free energy of the system, respectively. The free energy of the system is then expanded about the homogeneous reference state ({\it i.e.} the solid solution) in terms of the $\{c_{i\alpha}\}$. This Landau-type series expansion is written
\begin{align}
    \Omega^{(1)}[\{c_{i\alpha}\}] &= \Omega^{(1)}[\{c_{\alpha}\}] + \sum_{i\alpha} \frac{\partial \Omega^{(1)}}{\partial c_{i\alpha}} \Big\vert_{\{c_{\alpha}\}} \Delta c_{i\alpha} \nonumber \\ 
    &+ \frac{1}{2} \sum_{i\alpha; j\alpha'} \frac{\partial^2 \Omega^{(1)}}{\partial c_{i\alpha} \partial c_{j\alpha'}} \Big\vert_{\{c_{\alpha}\}} \Delta c_{i\alpha}\Delta c_{j\alpha'} + \dots.
\label{eq:landau}   
\end{align} 
The site-wise chemical potentials present in Eq.~\ref{eq:free_energy} serve as Lagrange multipliers in the linear response theory, but their variation is not relevant to the underlying physics so terms involving these derivatives are dropped~\cite{khan_statistical_2016, woodgate_compositional_2022, woodgate_short-range_2023}. The symmetry of the solid solution at high temperature---and the requirement that any imposed fluctuation conserves the overall concentrations of each chemical species---ensures that the first-order term vanishes. The change in free energy, $\delta \Omega^{(1)}$ as a result of a fluctuation is therefore written (to second order) as
\begin{equation}
    \delta \Omega^{(1)} = \frac{1}{2} \sum_{i\alpha; j\alpha'} \Delta c_{i\alpha} [\beta^{-1} \, C_{\alpha\alpha'}^{-1} - S^{(2)}_{i\alpha, j\alpha'}] \Delta c_{j\alpha'},
    \label{eq:chemical_stability_real}
\end{equation}
where $C_{\alpha \alpha'}^{-1} = \frac{\delta_{\alpha \alpha'}}{c_\alpha}$ is associated with the entropic contributions, and the term $-\frac{\partial^2 \langle \Omega_\text{el} \rangle_0}{\partial c_{i\alpha} \partial c_{j\alpha'}} \equiv S^{(2)}_{i\alpha;j\alpha'}$ is the second-order concentration derivative of the average electronic energy of the disordered alloy. The evaluation of this term has been covered in depth in earlier works \cite{khan_statistical_2016, woodgate_compositional_2022} and for brevity we will omit discussion of it here.

It should be noted that $S^{(2)}_{i\alpha;j\alpha'}$ is evaluated in reciprocal space in our codes, and therefore the change in free energy of Eq.~\ref{eq:chemical_stability_real} is written accordingly as:
\begin{equation}
    \delta \Omega^{(1)} = \frac{1}{2} \sum_{\bf k} \sum_{\alpha, \alpha'} \Delta c_\alpha({\bf k}) [\beta^{-1} C^{-1}_{\alpha \alpha'} -S^{(2)}_{\alpha \alpha'}({\bf k})] \Delta c_{\alpha'}({\bf k}).
\label{eq:chemical_stability_reciprocal}
\end{equation}
The matrix in square brackets $[\beta^{-1} C^{-1}_{\alpha \alpha'} -S^{(2)}_{\alpha \alpha'}({\bf k})]$, referred to as the chemical stability matrix, is related to an estimate of the ASRO, $\Psi_{i\alpha;j\alpha'}$. (Eigenvalues of the chemical stability matrices for the compounds considered in this study are visualised in the { Supplementary Material, Figures 1-26}.) When searching for an disorder-order transition, we start with the high temperature solid solution, lower the temperature and look for the temperature at which the lowest lying eigenvalue of this matrix passes through zero for any $\mathbf{k}$-vector in the irreducible Brillouin Zone. When this eigenvalue passes through zero at some temperature $T_\text{ord}$ and wavevector $\mathbf{k}_\text{ord}$, we infer the presence of an disorder-order transition with chemical polarisation $\Delta c_\alpha$ given by the associated eigenvector. In this fashion we can predict both dominant ASRO and also the temperature at which the solid solution becomes unstable and a crystallographically ordered phase emerges.

{
\subsection{Vibrational Effects}
\label{sec:vibrational_methods}
The free energy of Eq.~\ref{eq:free_energy} does not include a vibrational term. Full inclusion of these effects within our linear response analysis would be highly complex, requiring consideration of the impact of both atomic short- and long-range order on the elastic constants and phonon spectra of a given alloy for arbitrary chemical orderings. However, once a given chemical ordering has been identified (for example, for the L1$_0$ ordering identified for FeNi in Sec.~\ref{sec:feni}) it is possible to estimate the importance of vibrational effects in the following way.

When the change in free energy, Eq.~\ref{eq:chemical_stability_reciprocal}, is considered for an L1$_0$ ordering in binary FeNi, we only need consider the wavevector $\mathbf{k}=(0,0,1)$. Further, for a binary system, the $2\times2$ chemical stability matrix reduces to a single number since $\Delta c_{A}((0,0,1)) = -\Delta c_{B}((0,0,1)) = \Delta \tilde{c}$.
Writing $c = c_A$, we have
\begin{equation}
    [\beta^{-1} C^{-1}_{\alpha \alpha'} -S^{(2)}_{\alpha \alpha'}((0,0,1))] \mapsto \frac{1}{\beta c(1-c)} -S^{(2)}((0,0,1)),
\end{equation}
where $S^{(2)} = S^{(2)}_{11} + S^{(2)}_{22} - 2 S^{(2)}_{12}$. 

To include vibrational effects, we then assume that the change in vibrational free energy between ordered and disordered phases can be approximated for small chemical fluctuations as depending quadratically on the atomic long range order, $\Delta \tilde{c}$, {\it i.e.} $\Delta F_\text{vib}(T) = A(T)(\Delta \tilde{c})^2$. (Such quadratic dependence is justified by, {\it e.g.} Fig.~3 of Ref.~\cite{tian_density_2019}.) For the L1$_0$ ordering in FeNi, we estimate the temperature dependent parameter $A(T)$ by considering the difference in vibrational free energy between the fully ordered L1$_0$ phase and disordered A1 phase, writing 
\begin{equation}
    A(T) = F^\textrm{L1$_0$}_\textrm{vib}(T) - F^\textrm{A1}_\textrm{vib}(T),
    \label{eq:vibrational_contribution}
\end{equation}
where the vibrational free energies are evaluated within a Debye model as proposed by Moruzzi, Janak, and Schwarz~\cite{moruzzi_calculated_1988}.

The only free parameters in the above model are the Debye temperatures of the A1 and L$1_0$ phases. We obtain Debye temperatures for both phases within the approximation of Anderson~\cite{anderson_simplified_1963}, which is based on the density, bulk modulus, and shear modulus of a given phase. The bulk and shear modulus of each phase can be calculated from its elastic constants, which we obtain using CASTEP~\cite{clark_first_2005}. Our fitted elastic constants, bulk and shear moduli, and Debye temperatures are { given in Supplementary Table 3.}

Having obtained the bulk and shear moduli for both phases and, correspondingly, the Debye temperatures, we can use Eq.~\ref{eq:vibrational_contribution} to quantify the impact of vibrational effects on the predicted chemical ordering temperature. The (now $1\times 1$) chemical stability matrix is written
\begin{equation}
    \Psi^{-1}((0,0,1))) = \frac{1}{\beta c(1-c)} -S^{(2)}((0,0,1)) + A(T).
\end{equation}
From left to right, the three terms on the right hand side of this equation indicate contributions from entropic, internal energy, and vibrational terms, respectively. At a temperature of 500~K, our calculated entropic contribution is 172.3~meV/atom, $S^{(2)}((0,0,1))$ is 174.9~meV/atom, and our vibrational contribution is 18.4~meV/atom. Once the vibrational contribution to the free energy is included, our predicted L1$_0$ ordering temperature for FeNi is reduced from 508~K (without vibrational effects) to 449~K (with vibrational effects).
}

\subsection{Pairwise Atomistic Model}
\label{sec:atomistic}

To explore the alloy phase space further than is possible with the above linear response analysis, we map the concentration derivatives of the internal energy of the alloy, $S^{(2)}_{\alpha\alpha'}(\mathbf{k})$, to a pairwise real-space interaction. This real-space model, referred to as the Bragg-Williams model~\cite{bragg_effect_1934, bragg_effect_1935}, is lattice-based and has a Hamiltonian of the form
\begin{equation}
    H = \frac{1}{2}\sum_{i \alpha; j\alpha'} V_{i\alpha; j\alpha'} \xi_{i \alpha} \xi_{j \alpha'}.
    \label{eq:b-w2}
\end{equation}
The effective pairwise interactions, $V_{i\alpha; j\alpha'}$, are recovered from $S^{(2)}_{\alpha\alpha'}(\mathbf{k})$ by means of a inverse Fourier transform. The mapping from reciprocal-space to real-space and fixing of the gauge degree of freedom on the $V_{i\alpha; j\alpha'}$ is specified in earlier works~\cite{khan_statistical_2016, woodgate_compositional_2022, woodgate_short-range_2023}. These interactions are assumed to be isotropic and the sum in Eq.~\ref{eq:b-w} is then taken as a sum over coordination shells, {\it i.e.} first-nearest neighbours, second-nearest neighbours, \textit{etc.}

{ The pairwise form of the internal energy given in Eq.~\ref{eq:b-w2} can be thought of as related to a two-body term in a cluster expansion of the internal energy of the alloy. However, there is an important distinction to be made here: a cluster expansion typically involves fitting an expansion including both two-body and higher order terms to a large dataset of atomic configurations and associated DFT energies. By contrast, the $S^{(2)}$ theory requires only a single self-consistent KKR-CPA calculation to be performed to describe the average electronic structure and associated internal energy of the disordered alloy. The derivatives of this internal energy and subsequent real-space pairwise form of the internal energy are obtained via a perturbative analysis of the KKR-CPA effective medium. We note that, because the current {\it ab initio} framework evaluates a pair correlation function, a pair interaction is obtained. In principle, however, the methodology is extensible to evaluation of higher-order correlations and the corresponding many-body interactions, akin to a cluster expansion. We suggest that one approach which could be followed to obtain these would be analogous to that used in Refs.~\cite{mendive-tapia_theory_2017} and \cite{mendive_tapia_ab_2020}, which obtained higher-order interactions in a magnetic setting within a similar first-principles framework to that of the present work.}

\subsection{Monte Carlo Simulations}
\label{sec:monte_carlo}

To investigate the phase behaviour of these systems with the above atomistic model, we perform simulated annealing using the Metropolis Monte-Carlo algorithm with Kawasaki dynamics~\cite{landau_guide_2014}. These dynamics naturally conserve overall concentrations of each chemical species by permitting only swaps of pairs of atoms.

The algorithm starts by initialising the occupancies of each lattice site randomly, with the only constraint being the overall number of atoms of each chemical species, specified by the overall concentrations. A pair of atomic sites (not necessarily nearest-neighbours) is selected at random, and the change in internal energy $\Delta H$ resulting from swapping the site occupancies is calculated. If the change in energy is negative ($\Delta H < 0$) the swap is accepted unconditionally, while if the change is positive ($\Delta H > 0$) the swap is accepted with acceptance probability $e^{-\beta \Delta H}$. To collect the configurational contribution to the specific heat capacity (SHC) of the system, we use the fluctuation-dissipation theorem~\cite{allen_computer_2017} and the expression
\begin{equation}
    C = \frac{1}{k_b T^2} \left( \langle E^2 \rangle - \langle E \rangle^2 \right).
\end{equation}

To quantify the emergent ASRO in these simulations, we use the conditional probability of finding one species neighbouring another, writing $P^{pq}_n$ to denote the conditional probability of an atom of type $q$ neighbouring an atom of type $p$ on coordination shell $n$. In the high-temperature limit, these tend to the value $c_q$, {\it i.e.} the overall concentration of species $q$. Divergence from this homogeneous value is indicative of emergent atomic short- and long-range order.

{
\subsection{Magnetic critical (Curie) temperature}
\label{sec:TC_theory}
One of the most fundamental properties of of a ferromagnetic material is its magnetic critical temperature, $T_C$, the temperature at which its magnetic susceptibility diverges and below which magnetic order is established. For a ferromagnet, this quantity is known as the Curie temperature. In this work, we compute the magnetic critical temperatures of the phases studied in this work using the disordered local moment (DLM) description of magnetic disorder at finite temperature~\cite{pindor_disordered_1983, staunton_disordered_1984, gyorffy_first-principles_1985}.

The computational approach for using DLM calculations to obtain an estimate of the magnetic critical temperature of a system is described in detail in other works, and the reader is referred to those for full details of the underlying theory~\cite{gyorffy_first-principles_1985, patrick_rare-earthtransition-metal_2017, nawa_temperature-dependent_2020, patrick_calculating_2018}. In essence, for a given value of $\beta = 1/(k_B T)$, the approach expresses the magnetisation of a system, $m_{i\alpha}$ in terms of self-consistently determined Weiss fields, $h_{i\alpha}$, writing
\begin{equation}
m_{i\alpha} = L(\beta h_{i\alpha}),
\label{eq:langevin}
\end{equation}
where $L(x)$ is the Langevin function, defined as
\begin{equation}
L(x) = \coth(x) - \frac{1}{x},
\end{equation}
which reduces to $L(x) \approx x/3$ in the limit of small $x$. Therefore, in the limit of small $\beta h_{i\alpha}$, Eq.~\ref{eq:langevin} can be linearised to obtain an estimate of the Curie temperature.

In the case of a pure material with a single non-equivalent lattice site in the unit cell ($m_{i\alpha} = m$, $h_{i\alpha} = h$) the linearised equation is written~\cite{gyorffy_first-principles_1985}.
\begin{equation}
m \approx \frac{1}{3} \beta h. \label{eq:weiss}
\end{equation}
For small $m$ is is reasonable to suppose that the Weiss field is linear in $m$ and it is therefore possible to write
\begin{equation}
h = J m. \label{eq:weiss2}
\end{equation}
Given a few small values of $\beta h$, it is possible to perform a linear fit to obtain $J$, then the critical temperature is found as $T_C = J/(3k_B)$.

For the atomically ordered L1$_0$ phase of FeNi, where there are two non-equivalent lattice sites in the unit cell, the same linearisation can be performed but this time a matrix equation is obtained. A more detailed discussion of this procedure can be found in Ref.~\cite{nawa_temperature-dependent_2020}. The Weiss fields for the two non-equivalent Fe sites are written 
\begin{equation}
\begin{pmatrix} 
h_{\textrm{1a}} \\ h_{\textrm{1d}} 
\end{pmatrix}
\approx
\begin{pmatrix} 
J_{\textrm{1a-1a}} & J_{\textrm{1a-1d}} \\
J_{\textrm{1a-1d}} & J_{\textrm{1d-1d}}
\end{pmatrix}
\begin{pmatrix} 
m_{\textrm{1a}} \\ m_{\textrm{1d}} 
\end{pmatrix},
\end{equation}
where $h_{\textrm{1a}}$ is the Weiss field on the 1a site, and all other symbols are natural generalisations of the symbols of Equations~\ref{eq:weiss} and \ref{eq:weiss2} to the case of multiple sublattices. The critical temperature is then obtained in the limit of small $\beta h_i$ by solving the eigenvalue problem
\begin{equation}
\begin{pmatrix} 
h_{\textrm{1a}} \\ h_{\textrm{1d}} 
\end{pmatrix}
= \frac{\beta}{3}
\begin{pmatrix} 
J_{\textrm{1a-1a}} & J_{\textrm{1a-1d}} \\
J_{\textrm{1a-1d}} & J_{\textrm{1d-1d}}
\end{pmatrix}
\begin{pmatrix} 
h_{\textrm{1a}} \\ h_{\textrm{1d}} 
\end{pmatrix},
\end{equation}
where the smallest eigenvalue of the matrix
\begin{equation}
\frac{1}{3k_B}
\begin{pmatrix} 
J_{\textrm{1a-1a}} & J_{\textrm{1a-1d}} \\
J_{\textrm{1a-1d}} & J_{\textrm{1d-1d}}
\end{pmatrix}
\end{equation}
corresponds to $T_C$. The eigenvector associated with this eigenvalue then informs us of the relevant ordering strengths of different chemical species and/or magnetic sublattices~\cite{patrick_rare-earthtransition-metal_2017}.

The approach outlined can also be generalised to three or more magnetic sublattices, and to multiple magnetic species per lattice site, as is necessary for the partially and fully ordered crystal structures considered in this work.
}

\subsection{Evaluating magnetocrystalline anisotropy \emph{ab initio}}
\label{sec:MCA_theory}

The magnetocrystalline anisotropy of a system describes the change in energy of the system due a rotation of the magnetisation direction. For a ferromagnet with magnetisation direction $\hat{\mathbf{n}}$, the MCA can be expressed as
\begin{equation}
    K = \sum_\gamma K_\gamma g_\gamma (\hat{\mathbf{n}}),
\end{equation}
where $K_\gamma$ denote MCA coefficients, and $g_\gamma$ denote spherical harmonics fixing the orientation of $\hat{\mathbf{n}}$ with respect to the crystal axes. These spherical harmonics must respect the symmetry of the underlying crystal structure. It is convenient to represent $\hat{\mathbf{n}}$ in spherical polar coordinates by $\hat{\mathbf{n}} = (\sin\theta\cos\phi, \sin\theta\sin\phi, \cos\theta)$, where $\theta$ and $\phi$ denote polar and azimuthal angles respectively in a coordinate frame specified by the crystal axes. 

In a uniaxial ferromagnet, the MCA is well-approximated by
\begin{equation}
    K = K_1 \sin^2 \theta
\end{equation}
and the total energy of the system can be written as
\begin{equation}
    {\cal{E}}(\hat{\mathbf{n}}) = {\cal{E}}_\text{iso} + K_1 \sin^2 \theta,
    \label{eq:uniax_e}
\end{equation}
where $\cal{E}_\text{iso}$ represents the isotropic portion of the energy. The MCA coefficient $K_1$ can be calculated from derivatives of this energy with respect to magnetisation directions. Given an energy of the form in Eq.\ref{eq:uniax_e}, we have that
\begin{equation}
    \frac{\partial {\cal{E}}}{\partial \theta} = K_1 \sin 2\theta.
    \label{eq:ederiv}
\end{equation}
The magnetocrystalline anisotropy difference between the system magnetised in the $\hat{\mathbf{z}}$ and  $\hat{\mathbf{x}}$ directions is the sum, 
\begin{equation}
    \Delta {\cal{E}} = {\cal{E}}([1,0,0]) - {\cal{E}}([0,0,1]) = K_1.
\end{equation}
Evaluation of the derivative, $\frac{\partial \cal{E}}{\partial \theta}$ at $\theta = \frac{\pi}{4}$ gives $K_1$ directly, enabling us to obtain the MCA in a one-shot calculation. Evaluation at other angles enables extraction of coefficients of higher-order anisotropy coefficients, if desired.

Using the MARMOT code~\cite{patrick_marmot_2022}, which works with a fully relativistic formulation of density functional theory (DFT), these torques can be evaluated {\it ab initio}. A discussion of the relevant formulae for the torque in terms of key quantities of multiple scattering theory can be found in Refs.~\cite{staunton_temperature_2006} and \cite{woodgate_revisiting_2023}.

\subsection{Computational Details}
\label{sec:computational_details}

Throughout, the all-electron HUTSEPOT code~\cite{hoffmann_magnetic_2020} is used to generate self-consistent, one electron potentials within the KKR formulation of density functional theory (DFT)~\cite{martin_electronic_2004, faulkner_multiple_2018}. Chemical disorder is described within the coherent potential approximation (CPA)~\cite{faulkner_calculating_1980, johnson_total-energy_1990}, while the magnetic disorder of the paramegnetic state is modelled within the disordered local moment (DLM) picture~\cite{pindor_disordered_1983, staunton_disordered_1984, gyorffy_first-principles_1985}. We perform spin-polarised, scalar-relativistic calculations within the atomic sphere approximation (ASA)~\cite{stocks_complete_1978} with an angular momentum cutoff of $l_\text{max} = 3$ for basis set expansions, a $20\times20\times20$ $\mathbf{k}$-point mesh for integrals over the Brillouin zone, and a 24 point semi-circular Gauss-Legendre grid in the complex plane to integrate over valence energies. We use the local density approximation (LDA) and the exchange-correlation functional is that of Perdew-Wang~\cite{perdew_accurate_1992}. The linear response results are obtained from an computational implementation of the theory described in Section~\ref{sec:chemical_ordering} and  discussed in detail in earlier works~\cite{khan_statistical_2016, woodgate_compositional_2022, woodgate_short-range_2023}.

Lattice parameters for the alloyed systems are obtained by calculating an fcc lattice parameter based on a weighted average of atomic volumes for the pure elements, taken from Ref.~\cite{hermann_crystallography_2011}. Full details are provided in { Supplementary Tables 1 and 2}. For example, for the binary FeNi system, this results in a cubic fcc lattice parameter of $a=3.57$~\AA, while for the ternary Fe$_4$Ni$_3$Pt system, this results in a cubic fcc lattice parameter of $a=3.62$~\AA. Use of the experimentally determined atomic volumes, which will be slightly larger than LDA estimates, naturally incorporating the effects of thermal expansion in our modelling. Throughout, we fix $c/a=1$, which is justified for these FeNi-based systems, as L1$_0$ FeNi is known from both experimental and computational studies to have minimal tetragonal distortion to the lattice, with $c/a$ ratio being between 1.00026 (experimental determination, Ref.~\cite{lewis_inspired_2014}) and 1.0048 (GGA estimate, Ref.~\cite{izardar_interplay_2020}). In previous studies, we have found that our $S^{(2)}$ calculations are only weakly sensitive to the choice of lattice parameter around the experimental value~\cite{woodgate_competition_2024}.

Magnetocrystalline anisotropy energies { and magnetic critical temperatures} are evaluated using the MARMOT code~\cite{patrick_marmot_2022}, which takes in { ferromagnetic} HUTSEPOT potentials { (frozen potential approximation)} as a starting point for its evaluation of the { Weiss fields and } magnetic torques as described in {Sections~\ref{sec:TC_theory} and} \ref{sec:MCA_theory}. Rather than include the effects of spin-orbit coupling in an approximate way, by adding another term to the scalar-relativistic Hamiltonian, MARMOT works in a fully relativistic setting in which spin-orbit effects are included naturally. We use parameters consistent with the HUTSEPOT calculations, and other relevant parameters are left at their MARMOT default values. This includes angular sampling of the CPA integral of $240\times40$ and an adaptive meshing scheme for Brillouin zone integrations.

{
For calculations of the elastic constants, bulk and shear moduli, and associated Debye temperatures of the A1 and L1$_0$ phases of FeNi, used for assessing the impact of lattice vibrations on predicted chemical ordering temperatures, we use CASTEP~\cite{clark_first_2005}, which is a plane-wave DFT package. All calculations were performed in the ferromagnetic state. The L1$_0$ structure is described by its two-atom bct unit cell, while results for the A1 phase are obtained by averaging quantities across three 32-atom special quasirandom structures~\cite{zunger_special_1990} (SQSs) as implemented in the ICET package~\cite{angqvist_icet_2019}. We use the Generalised Gradient Approximation (GGA) and the exchange-correlation functional of Perdew, Burke, and Ernzerhof (PBE)~\cite{perdew_generalized_1996}. A plane-wave cutoff energy of 600~eV was used in conjunction with an electronic self-consistency criterion of $1\times10^{-6}$~eV. For the two-atom L1$_0$ structure, a $12\times 12\times 12$ Monkhorst--Pack grid~\cite{monkhorst_special_1976} was used for Brillouin zone integration, while for the 32-atom supercells a $6\times 6\times 6$ grid was used. Structures were initially relaxed to obtain optimised geometries~\cite{payne_iterative_1992, pfrommer_relaxation_1997, byrd_representations_1994}, with an atomic force tolerance of 0.03~eV\AA$^{-1}$. The average optimised fcc lattice constant for the A1 phase is 3.570~\AA, while for the L1$_0$ phase, $a=b=3.556$~\AA and $c=3.577$~\AA, corresponding to $c/a=1.006$. The optimised cells were then sheared according to the scheme outlined in Ref.~\cite{walker_effect_2012}, from which the elastic constants, bulk modulus, and shear modulus, can be fitted.
}

\section*{Data Availability}
\label{sec:data_availability}
The data produced for this study are available through Zenodo via the DOI \href{https://zenodo.org/doi/10.5281/zenodo.10425681}{10.5281/zenodo.10425681}. Specific questions should be directed to the corresponding author(s).

\section*{Code Availability}
\label{sec:code_availability}
The all-electron HUTSEPOT code~\cite{hoffmann_magnetic_2020} used for constructing the self-consistent one-electron potentials of DFT is available at \href{https://hutsepot.jku.at/}{hutsepot.jku.at}. The code implementing the $S^{(2)}$ theory for multicomponent alloys~\cite{khan_statistical_2016, woodgate_modelling_2024} is available from the authors on reasonable request. The code for performing lattice-based Monte Carlo simulations to study alloy phase behaviour using the obtained atom-atom interchange parameters~\cite{woodgate_compositional_2022} is available via the DOI \href{https://doi.org/10.5281/zenodo.10379949}{10.5281/zenodo.10379949}. The MARMOT code~\cite{patrick_marmot_2022}, used for evaluating { magnetic critical temperatures and } magnetocrystalline anisotropy energies { within the disordered local moment (DLM) picture}, is available at \href{https://warwick.ac.uk/marmotcode/}{warwick.ac.uk/marmotcode}. {CASTEP (for structural optimisation, elastic constants, and bulk moduli) is available at \href{http://www.castep.org}{www.castep.org}.}

\begin{acknowledgments}
The authors would like to express their gratitude to Dr Christopher E. Patrick (Department of Materials, University of Oxford) for fruitful discussions, software support, and feedback on the draft version of this manuscript.

This work was supported by the UK Engineering and Physical Sciences Research Council, Grant No. EP/W021331/1, by the U.S. Department of Energy, Office of Basic Energy Sciences under Award Number DE SC0022168 (for atomic insight) and by the U.S. National Science Foundation under Award ID 2118164 (for advanced manufacturing aspects). C.D.W. was also supported by a studentship within the UK Engineering and Physical Sciences Research Council-supported Centre for Doctoral Training in Modelling of Heterogeneous Systems, Grant No. EP/S022848/1. Computing facilities were provided by the Scientific Computing Research Technology Platform (SCRTP) of the University of Warwick.
\end{acknowledgments}

\section*{Author Contributions}
\label{sec:author_contributions}
C.D.W., L.H.L., and J.B.S. jointly conceived of the approach. J.B.S. led development of the MARMOT package used for the magnetocrystalline anisotropy calculations and also developed the  code for the ASRO linear response calculations. The electronic structure calculations, linear response analysis, and evaluation of magnetocrystalline anisotropy energies were performed by C.D.W. and J.B.S. { Calculations of magnetic critical temperatures were performed by C.D.W.} Atomistic simulations and subsequent analysis were performed by C.D.W. {Calculations of elastic constants, bulk moduli, Debye temperatures, and vibrational free energies were performed by C.D.W.} The first draft of the manuscript was written by C.D.W. and subsequently input was received from all authors. J.B.S. and L.H.L. supervised the project and led funding acquisition.

\section*{Competing Interests}
\label{sec:competing_interests}
{
J.B.S. acts as an associate editor for npj Computational Materials. The other authors declare no competing interests.
}

\end{document}